\documentclass{aa}

\usepackage{graphicx}
\usepackage{epsfig}
\usepackage{natbib}
\usepackage{pifont}

\usepackage[varg]{txfonts}
\usepackage[font=footnotesize,labelfont={bf,sc,small},figurename=Fig.]{caption}
\usepackage[colorlinks,allcolors=blue]{hyperref}
\usepackage{subfig}
\usepackage{rotate}
\usepackage{appendix}

\usepackage{varioref}

\def\PLUTO{{\sc pluto}}

\newcommand\rs[1]{_\mathrm{#1}}
\newcommand{\casa}{Cas~A}

\begin{document} 

%%\title{Reverse shock of Cassiopeia A reveals past interaction\\ with massive circumstellar shell}
\title{Evidence for past interaction with an asymmetric circumstellar
shell in the young SNR Cassiopeia A}

\author{S.\ Orlando\inst{1}
   \and A.\ Wongwathanarat\inst{2}
   \and H.-T.\ Janka\inst{2}
   \and M.\ Miceli\inst{3,1}
   \and S.\ Nagataki\inst{4,5} 
   \and M.\ Ono\inst{4,5} \and \\
   F.\ Bocchino\inst{1}
   \and J.\ Vink\inst{6,7,8}
   \and D.\ Milisavljevic\inst{9}
   \and D.J.\ Patnaude\inst{10}
   \and G.\ Peres\inst{3,1}
}

\offprints{S. Orlando}

\institute{INAF -- Osservatorio Astronomico di Palermo, Piazza del Parlamento 1, I-90134 Palermo, Italy\\ 
\email{salvatore.orlando@inaf.it}
\and Max-Planck-Institut f\"ur Astrophysik, Karl-Schwarzschild-Str. 1, D-85748 Garching, Germany
\and Dip. di Fisica e Chimica, Universit\`a degli Studi di Palermo, Piazza del Parlamento 1, 90134 Palermo, Italy
\and Astrophysical Big Bang Laboratory, RIKEN Cluster for Pioneering Research, 2-1 Hirosawa, Wako, Saitama 351-0198, Japan
\and RIKEN Interdisciplinary Theoretical \& Mathematical Science Program (iTHEMS), 2-1 Hirosawa, Wako, Saitama 351-0198, Japan
\and Anton Pannekoek Institute for Astronomy, University of Amsterdam, Science Park 904, 1098 XH Amsterdam, The
Netherlands
\and GRAPPA, University of Amsterdam, Science Park 904, 1098 XH Amsterdam, The Netherlands
\and SRON, Netherlands Institute for Space Research, Utrech, The Netherlands
\and Department of Physics and Astronomy, Purdue University, 525 Northwestern Avenue, West Lafayette, IN 47907, USA
\and Smithsonian Astrophysical Observatory, 60 Garden Street, Cambridge, MA 02138, USA
  }

\date{Received date / Accepted date}

\abstract
%Context
{Observations of the supernova remnant (SNR) Cassiopeia A (\casa)
show significant asymmetries in the reverse shock that cannot be
explained by models describing a remnant expanding through a
spherically symmetric wind of the progenitor star.}
%aims
{We investigate whether a past interaction of \casa\ with a massive
asymmetric shell of the circumstellar medium can account for
the observed asymmetries of the reverse shock.}
%Methods
{We performed three-dimensional (3D) (magneto)-hydrodynamic simulations
that describe the remnant evolution from the SN explosion to its
interaction with a massive circumstellar shell. The initial conditions
(soon after the shock breakout at the stellar surface) are provided
by a 3D neutrino-driven SN model whose morphology closely resembles
\casa\ (\citealt{2017ApJ...842...13W}) and the SNR simulations
cover $\approx 2000$~years of evolution. We explored the parameter
space of the shell, searching for a set of parameters able to produce
an inward-moving reverse shock in the western hemisphere of
the remnant at the age of $\approx 350$~years, analogous to that
observed in \casa.}
%Results
{The interaction of the remnant with the shell can produce
asymmetries resembling those observed in the reverse shock if the
shell was asymmetric with the densest portion in the (blueshifted)
nearside to the northwest (NW). According to our favorite model,
the shell was thin (thickness $\sigma \approx 0.02$~pc) with a
radius $r\rs{sh} \approx 1.5$~pc from the center of the explosion.
The reverse shock shows the following asymmetries at the age of
\casa: i) it moves inward in the observer frame in the NW region,
while it moves outward in most other regions; ii) the geometric
center of the reverse shock is offset to the NW by $\approx 0.1$~pc
from the geometric center of the forward shock; iii) the reverse
shock in the NW region has enhanced nonthermal emission because,
there, the ejecta enter the reverse shock with a higher relative
velocity (between $4000$ and $7000$~km~s$^{-1}$) than in other
regions (below $2000$~km~s$^{-1}$).}
%Conclusions
{The large-scale asymmetries observed in the reverse shock of \casa\
can be interpreted as signatures of the interaction of the remnant
with an asymmetric dense circumstellar shell that occurred between
$\approx 180$ and $\approx 240$~years after the SN event. We suggest
that the shell was, most likely, the result
of a massive eruption from the progenitor star that occurred between
$10^4$ and $10^5$~years prior to core-collapse. We estimate a total
mass of the shell of the order of $2\,M_{\odot}$.}


\titlerunning{Cassiopeia A reveals past interaction with circumstellar shell}
\authorrunning{S. Orlando et~al.}

\maketitle

\section{Introduction}
\label{sec:intro}

Cassiopeia A (in the following \casa) is one of the best studied
supernova remnants (SNRs) of our Galaxy. Its relative youth (with
an age of $\approx 350$~years; \citealt{2006ApJ...645..283F}) and
proximity (at a distance of $\approx 3.4$~kpc;
\citealt{1995ApJ...440..706R}) make this remnant
an ideal target to study the structure and chemical composition of
the stellar material ejected by a supernova (SN). The analysis of
multi-wavelength observations allowed some authors to reconstruct
in great details its three-dimensional (3D) structure (e.g.,
\citealt{2010ApJ...725.2038D, 2013ApJ...772..134M, 2015Sci...347..526M,
2014Natur.506..339G, 2017ApJ...834...19G}). Several lines of evidence
suggest that the morphology and expansion rate of \casa\ are
consistent with a remnant mainly expanding through a spherically
symmetric wind of the progenitor star (e.g.,
\citealt{2014ApJ...789....7L}). Thus, the vast majority of anisotropies
observed in the remnant morphology most likely reflect asymmetries
left from the earliest phases of the SN explosion. This makes \casa\
a very attractive laboratory to link the physical, chemical and
morphological properties of a SNR to the processes at work during
the complex phases of the SN.

First attempts to link \casa\ to its parent SN were very successful
and have shown that the bulk of asymmetries observed in the remnant
are intrinsic to the explosion (\citealt{2016ApJ...822...22O}), and
the extended shock-heated Fe-rich regions evident in the main shell
originate from large-scale asymmetries that developed from stochastic
processes (e.g., convective overturn and the standing accretion
shock instability; SASI) during the first seconds of the SN blast
wave (\citealt{2017ApJ...842...13W}). More recently,
\cite{2021A&A...645A..66O} (in the following Paper I) have extended
the evolution of the neutrino-driven core-collapse SN presented in
\cite{2017ApJ...842...13W} till the age of 2000 years with the aim
to explore how and to which extent the remnant keeps memory of
post-explosion anisotropies imprinted to the ejecta by the asymmetric
explosion mechanism. Comparing the model results for the SNR at
$\approx 350$~years with observations shows that the main asymmetries
and features observed in the ejecta distribution of \casa\ result
from the interaction of the post-explosion large-scale anisotropies
in the ejecta with the reverse shock.

The above models, however, do not explain one of the most
intriguing aspects of the \casa\ structure, as evidenced by the
analysis of the position and velocity of the forward and reverse
shocks. Observations in different wavelength bands indicate a forward
shock expanding with a velocity around $\approx 5500$~km~s$^{-1}$
along the whole remnant outline (e.g.,~\citealt{2003ApJ...589..818D,
2009ApJ...697..535P, 2019sros.confE..32F, 2022ApJ...929...57V}) and
a reverse shock moving outward with velocity ranging between
$2000$~km~s$^{-1}$ and $4000$~km~s$^{-1}$ in the eastern and northern
hemisphere of the remnant (e.g.,~\citealt{2018ApJ...853...46S,
2019sros.confE..32F, 2022ApJ...929...57V}). These velocities are
somehow consistent with the remnant expanding through a spherically
symmetric wind of the progenitor star and, in fact, are well
reproduced by the models (e.g., \citealt{2016ApJ...822...22O} and
Paper I). The observations, however, suggest that the reverse shock
in the southern and western quadrants of \casa\ is stationary or
is even moving inward in the observer frame toward the center of
the explosion (e.g., \citealt{1995ApJ...441..307A, 1996ApJ...466..309K,
2004ApJ...613..343D, 2004ApJ...614..727M, 2008ApJ...686.1094H,
2018ApJ...853...46S}) at odds with the model predictions (e.g.,
\citealt{2020pesr.book.....V, 2022ApJ...929...57V}; see also Paper
I). In addition, observations of \casa\ show an offset of $\approx
0.2$~pc (at the distance of $3.4$~kpc) between the geometric center
of the reverse shock and that of forward shock
(\citealt{2001ApJ...552L..39G}) that cannot be reproduced by the
models (see Paper I).

\begin{figure}[!t]
  \begin{center}
    \leavevmode
        \epsfig{file=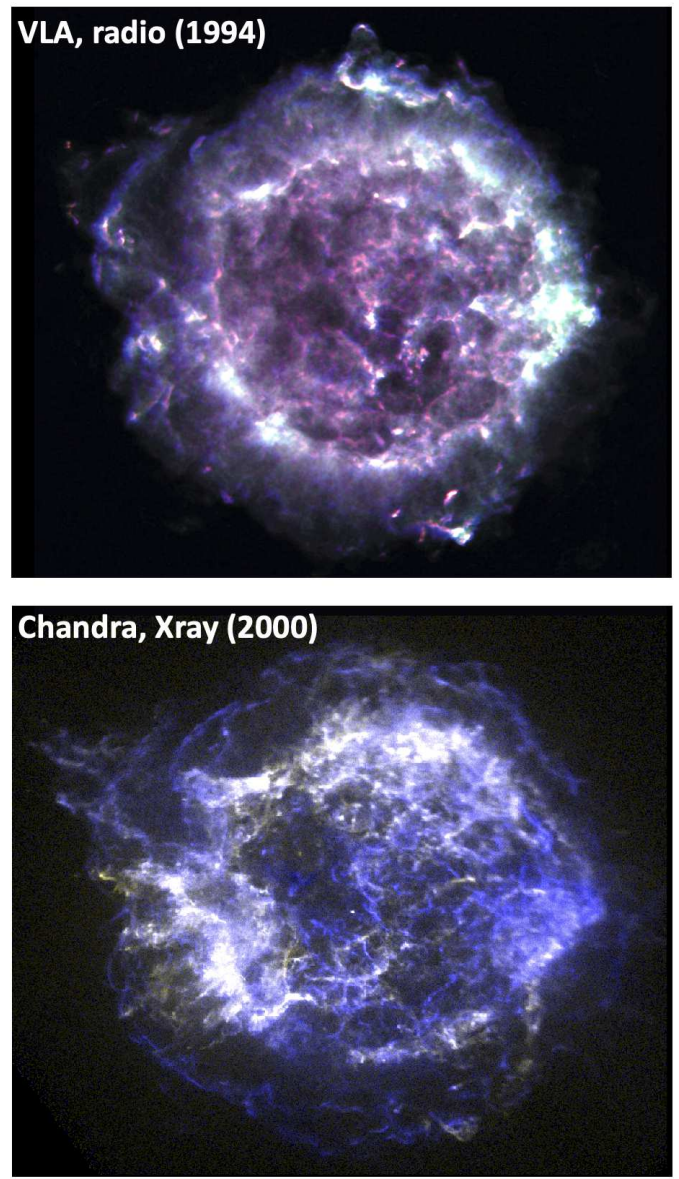, width=9cm}
	\caption{Upper panel: radio observation of \casa\ collected with
	the Very Large Array telescope in 1994 at the frequencies 1.4
	GHz (L band), 5.0 GHz (C band) and 8.4 GHz (X band), which
	highlights the emission dominated by synchrotron radiation
	(Credit: L. Rudnick, T. Delaney, J. Keohane, B. Koralesky
	and T. Rector; NRAO/AUI/NSF). Lower panel: X-ray observation
	of \casa\ collected with Chandra in 2000, in sqrt scale,
	showing the emission in the $[0.5, 2]$~keV (white) and $[2,
	7]$~keV (blue) bands.}
  \label{fig_casa}
\end{center} \end{figure}

The above results are even more puzzling by looking at the X-ray
synchrotron emission associated with the forward and reverse shocks.
\cite{2008ApJ...686.1094H} have shown that the reverse shock radiation
is limited to a thin spherical shell partially visible mainly in
the western hemisphere and shifted toward the west with respect to
the remnant outline (thus again suggesting an offset between the
centers of the reverse and forward shocks). A similar conclusion
was reached by analyzing radio observations of \casa\ (see
upper panel of Fig.~\ref{fig_casa}), which show that the forward
and reverse shocks are much closer to each other and the radio
emission is higher in the western than in the eastern hemisphere
(\citealt{2018A&A...612A.110A}). \cite{2008ApJ...686.1094H} have
proposed that the high synchrotron emission in the western hemisphere
is due to a locally higher reverse shock velocity in the ejecta
rest frame ($\approx 6000$~km~s$^{-1}$), so that, there, the reverse
shock is able to accelerate electrons to the energies needed to
emit X-ray synchrotron radiation (see lower panel of
Fig.~\ref{fig_casa}).

Some hints about the possible cause of the unexpected reverse shock
dynamics come from the evidence that isolated knots of ejecta show
a significant blue and redshift velocity asymmetry: ejecta traveling
toward the observer have, on average, lower velocities than ejecta
traveling away (e.g., \citealt{1995ApJ...440..706R, 2010ApJ...725.2038D,
2013ApJ...772..134M}). It is debated as to whether this is due to
the explosion dynamics (\citealt{2010ApJ...725.2038D, 2010ApJ...725.2059I})
or expansion into inhomogeneous circumstellar medium (CSM;
\citealt{1995ApJ...440..706R, 2013ApJ...772..134M}). Observations
of slow-moving shocked circumstellar clumps in the remnant, the
so-called ``quasi-stationary flocculi'' (QSFs), seem to favor the
latter scenario. In fact, these structures are, in large majority,
at blue shifted velocities (\citealt{1995ApJ...440..706R}), implying
that more CSM material is placed in the front than in the back of
\casa. This may suggest that the asymmetries associated with the
velocities of ejecta knots and, most likely, the evidently asymmetric
structure of the reverse shock in \casa\ may reflect the interaction
of the remnant with an inhomogeneous structure of the CSM.

Further support to the scenario of inhomogeneous CSM comes from
radio and X-ray observations which suggest that the remnant is
interacting with a density jump in the ambient medium (probably a
local molecular cloud) in the western hemisphere
(\citealt{1996ApJ...466..309K, 2006MNRAS.371..829S, 2012ApJ...746..130H,
2014ApJ...796..144K}). More recently, \cite{2018ApJ...853...46S}
have analyzed Chandra and NuSTAR observations of \casa\ (see
lower panel of Fig.~\ref{fig_casa}), identifying inward-moving
shocks in the observer frame located from a region close to the
compact central object (inside the mean reverse shock radius derived
by \citealt{2001ApJ...552L..39G}) to the maximum of the dense shell
brightness to the west (coincident with the reverse-shock circle).
The authors connected these shocks with the brightest features in
X-ray synchrotron radiation seen with NuSTAR. Since in spherical
symmetry, an inward-moving reverse shock is not consistent with
the dynamical age of a SNR as young as \casa, they proposed that
the inward moving shocks are reflected shocks caused by the interaction
of the blast wave with a molecular cloud with a density jump $> 5$.

The possibility that the remnant is interacting with molecular clouds is
reasonable and may explain some of the features of the reverse shock
(e.g., \citealt{2014ApJ...796..144K, 2018ApJ...853...46S}).
This interaction would imply a deceleration of the forward
shock that propagates through a denser medium and, possibly, an
indentation in the remnant outline (e.g., \citealt{2015SSRv..188..187S}).
However, both of these signatures of interaction are not clearly
visible in \casa: the forward shock shows similar velocities along
the remnant outline and both the forward and reverse shocks have
shapes that are roughly spherical, without any sign of interaction
with a molecular cloud. Furthermore, most of the molecular gas
detected lies in the foreground of \casa\ (e.g.,
\citealt{2004Natur.432..596K, 2005A&A...430..561W, 2009MNRAS.394.1307D,
2018ApJ...866..139K}) and would not have any effect on the propagation
of the forward and reverse shocks.

If, on one hand, an ongoing interaction of the remnant with a
molecular cloud seems not to be plausible to justify the asymmetries
observed in the reverse shock of \casa, it is possible, on the other
hand, that the remnant has encountered a dense shell of the
CSM in the past (e.g.,~\citealt{1996ApJ...466..866B}) and the
signatures of that interaction are now visible in the structure of
the reverse shock. In fact, massive stars are known to experience
episodic and intense mass loss events before going to SN. These
events may be related, for instance, with the activity of luminous
blue variable stars (LBVs; \citealt{1984IAUS..105..233C}) and
Wolf-Rayet stars (WR stars; \citealt{2007ApJ...657L.105F,
2007Natur.447..829P, 2008MNRAS.389..131P, 2020MNRAS.492.5897S}).
In these cases, after the explosion, the shock wave from the SN
travels through the wind of the progenitor and, at some point,
collides with the material of pre-SN mass loss events (see
\citealt{2014ARA&A..52..487S} for a recent review).  For instance,
strong indications of interaction with a circumstellar shell, likely
associated with wind residual, have been recently found in the Vela
SNR (\citealt{2021A&A...649A..56S}).

Observations of light echoes showed that \casa\ is the remnant of
a Type IIb SN (\citealt{2008Sci...320.1195K, 2011ApJ...732....3R}).
This implies that its progenitor star has shed almost all of its H
envelope (see also \citealt{1976ApJS...32..351K, 1978ApJ...219..931C})
before the core-collapse. Various hypotheses have been proposed to
explain how the progenitor star of \casa\ has lost its envelope:
via its own stellar wind (e.g., \citealt{2003ApJ...591..288H}), or
via binary interaction that involves mass transfer and, possibly,
a common-envelope phase (e.g., \citealt{1992ApJ...391..246P}), or
via interaction of the progenitor with the first SN of a binary
that removed its envelope (\citealt{2020MNRAS.499.1154H}). In any
case, the expanding remnant, at some point, should have encountered
and interacted with the gas of these pre-SN mass loss events. In
an early study, \cite{1989ApJ...344..332C} have suggested that
\casa\ interacted with a circumstellar shell in the past and
identified its bright ring with the shocked shell. A few years
later, this idea was further investigated by \cite{1996ApJ...466..866B}
through a 1D numerical model. These authors have proposed that the
blast wave of \casa\ traveled through an inhomogeneous CSM characterized
by a circumstellar shell resulted from the interaction of the slow
stellar wind in the red supergiant stage of the progenitor star
with the faster wind in the subsequent blue supergiant stage.
\cite{2018ApJ...866..139K} have interpreted the spatial distribution
of QSFs observed in \casa\ as evidence of a massive eruption from
the progenitor system to the west, which most likely occurred
$10^4-10^5$~years before the SN. Observations of the circumstellar
environment around \casa\ have also shown evidence of nebulosities
that have been interpreted to be the relics of the red supergiant
mass-loss material from \casa's progenitor (\citealt{2020ApJ...891..116W}).

Here, we investigate whether some of the large-scale asymmetries
in the reverse shock of \casa\ may reflect the past interaction of
the remnant with a dense shell of CSM, most likely the consequence
of an episodic mass loss from the progenitor massive star that
occurred in the latest phases of its evolution before collapse. To
this end, we reconsidered our model for describing the remnant of
a neutrino-driven core-collapse SN that reproduces the main features
of \casa\ (presented in Paper I), but added the description of an
asymmetric shell of CSM with which the remnant interacts within the
first 300 years of evolution. We performed an extensive
simulation campaign to explore the parameter space of the shell and
derived, from the models, the profiles of the forward and reverse
shock velocity versus the position angle in the plane of the sky
at the age of \casa. By comparing the profiles derived from the
models with those inferred from the observations
(\citealt{2022ApJ...929...57V}) we identified the models which
better than others reproduce the observations. Since no complete
survey of parameter space can possibly be done, we do not expect
an accurate match between models and observations and we cannot
exclude that shells with structure different from those explored
here can do a better job in matching the observations.

Indeed, we do not aim at deriving an accurate, unique, description
of the CSM. Our idealized shell model aims at showing that the main
large-scale asymmetries observed in the reverse shock of \casa\
(namely, the inward-moving reverse shock observed in the western
hemisphere, the offset between the geometric centers of the reverse
and forward shocks, and the evidence that the nonthermal emission
from the reverse shock is brighter in the western than in the eastern
region) can be naturally explained as the result of a past interaction
of the remnant with a circumstellar shell. The study is also
relevant for disentangling the effects from interior inhomogeneities
and asymmetries (produced soon after the core-collapse) from those
produced by the interaction of the remnant with an inhomogeneous
CSM. Future studies are expected to consider a structure
of the CSM derived self-consistently from the mass-loss history of
the progenitor system; in this way, the comparison between models
derived from different progenitors and observations of \casa\ would
be able to provide some hints on the nature and mass-loss history
of the stripped progenitor of \casa\ and, possibly, to shed light
on the question whether it was a single star or a member in a binary.

The paper is organized as follows. In Sect.~\ref{sec:model} we
describe the model setup; in Sect.~\ref{sec:results} we discuss the
results for the interaction of the SNR with the asymmetric dense
shell of CSM; and in Sect.~\ref{sec:conclusion} we summarize the
main results and draw our conclusions. In Appendix \ref{new_model},
we discuss an alternative model for the asymmetric circumstellar
shell.

\section{Problem description and numerical setup}
\label{sec:model}

We adopted the numerical setup presented in Paper I and which
describes the full development of the remnant of a neutrino-driven
SN, following its evolution for $\approx 2000$~years. The setup is
the result of the coupling between a model describing a SN explosion
with remarkable resemblance to basic properties of \casa\ (model
W15-2-cw-IIb; \citealt{2017ApJ...842...13W}) and hydrodynamic (HD)
and magneto-hydrodynamic (MHD) simulations that describe the formation
of the full-fledged SNR (e.g.,~\citealt{2016ApJ...822...22O}). The
3D SN simulation follows the evolution from about 15 milliseconds
after core bounce to the breakout of the shock wave at the stellar
surface at about 1 day after the core-collapse. Then, the output
of this simulation was used as initial condition for 3D simulations
which follow the transition from the early SN phase to the emerging
SNR and the subsequent expansion of the remnant through the wind
of the progenitor star. A thorough description of the setup can be
found in Paper I, while a summary of its main features is provided
in Sect.~\ref{sec:sn-snr}. In this paper, the setup was used to
investigate if some of the features observed in the reverse shock
of \casa\ can be interpreted as signatures of the interaction of
the remnant with an asymmetric shell of dense CSM material, most
likely erupted by the progenitor star before its collapse (see
Sect.~\ref{sec:csm}).

\subsection{Modeling the evolution from the SN to the SNR}
\label{sec:sn-snr}

The SN model is described in \cite{2017ApJ...842...13W} and considers
the collapse of an original $15\,M_{\odot}$ progenitor star (denoted
as s15s7b2 in \citealt{1995ApJS..101..181W} and W15 in
\citealt{2015A&A...577A..48W}) from which its H envelope has been
removed artificially (before the collapse) down to a rest of $\approx
0.3\,M_{\odot}$ (the modified stellar model is termed W15-IIb in
\citealt{2017ApJ...842...13W}). This is motivated by the evidence that
observations of light echoes suggest that \casa\ is the remnant of
a Type IIb SN (\citealt{2008Sci...320.1195K, 2011ApJ...732....3R}),
so that its progenitor star has shed almost all of its H envelope
before to go SN (see also \citealt{1976ApJS...32..351K,
1978ApJ...219..931C}). The model also considers a neutrino-energy
deposition able to power an explosion with an energy of $1.5\times
10^{51}$~erg $= 1.5$~bethe $= 1.5$~B (see \citealt{2013A&A...552A.126W,
2015A&A...577A..48W}). After the explosion an amount of $3.3\,M_{\odot}$
of stellar debris was ejected into the CSM. The main model parameters
are summarized in Table~\ref{tabmod}.

The SN model takes into account: the effects of gravity (both
self-gravity of the SN ejecta and the gravitational field of a
central point mass representing the neutron star that has formed
after core bounce at the center of the explosion); the fallback of
material on the neutron star; the Helmholtz equation of state
(\citealt{2000ApJS..126..501T}), which includes contributions from
blackbody radiation, ideal Boltzmann gases of a defined set of fully
ionized nuclei, and arbitrarily degenerate or relativistic electrons and
positrons. In addition, the model considers a small $\alpha$-network
to trace the products of explosive nucleosynthesis that took place
during the first seconds of the explosion (see
\citealt{2013A&A...552A.126W, 2015A&A...577A..48W}). This nuclear
reaction network includes 11 species: protons ($^{1}$H), $^{4}$He,
$^{12}$C, $^{16}$O, $^{20}$Ne, $^{24}$Mg, $^{28}$Si, $^{40}$Ca,
$^{44}$Ti, $^{56}$Ni, and an additional ``tracer nucleus'' $^{56}$X,
which represents Fe-group species synthesized in neutron-rich
environments as those found in neutrino-heated ejecta (see
\citealt{2017ApJ...842...13W} for details).

After the shock breakout, the SN model shows large-scale asymmetries
in the ejecta distribution. The most striking features are three
pronounced Ni-rich fingers that may correspond to the extended
shock-heated Fe-rich regions observed in \casa. These features
naturally developed from stochastic processes (e.g. convective
overturn and SASI) during the first second after core bounce
(\citealt{2017ApJ...842...13W}). These characteristics make the
adopted SN model most promising to describe a remnant with properties
similar to those observed in \casa\ (see also Paper I).

The output of model W15-2-cw-IIb at $\approx 20$~hours after the
core-collapse was used as initial conditions for the 3D HD and MHD
simulations which describe the long-term evolution ($\approx
2000$~years) of the blast wave and ejecta, from the shock breakout
to the expansion of the remnant through the CSM. In Paper I, we
analyzed three long-term simulations to evaluate the effects of
energy deposition from radioactive decay and the effects of an
ambient magnetic field by switching these effects either on or off.
Here, we reconsidered the models presented in Paper I and modified
the geometry and density distribution of the CSM to describe the
interaction of the remnant with a dense shell in the CSM.

Our simulations include: i) the effects of energy deposition
from the dominant radioactive decay chain $^{56}{\rm Ni}
\rightarrow\ ^{56}{\rm Co} \rightarrow\ ^{56}{\rm Fe}$, by adding
a source term for the internal energy which takes into account the
energy deposit which can be converted into heat (excluding neutrinos,
which are assumed to escape freely) and assuming local energy
deposition without radiative transfer\footnote{We note that our
simplified treatment of energy deposition by radioactive decay
assumes that all of the decay energy is deposited in the ejecta
without any $\gamma$-ray leakage from the inner part of the remnant.
Thus, our models are expected to overestimate the effects of decay
heating and, possibly, to overlook some features. Nevertheless, as
shown in Sect.~\ref{param_space}, although the effects of decay
heating are overestimated in our simulations, the remnant evolution
during its interaction with the shell is similar in models either
with or without these effects.} (\citealt{1999astro.ph..7015J,
2019ApJ...877..136F}); ii) the deviations from equilibrium of
ionization, calculated through the maximum ionization age in each
cell of the spatial domain (see \citealt{2015ApJ...810..168O});
iii) the deviations from electron-proton temperature equilibration,
calculated by assuming an almost instantaneous heating of electrons
at shock fronts up to $kT = 0.3$~keV (\citealt{2007ApJ...654L..69G})
and by implementing the effects of Coulomb collisions for the
calculation of ion and electron temperatures in the post-shock
plasma (\citealt{2015ApJ...810..168O}); and iv) the effects of
back-reaction of accelerated cosmic rays at shock fronts, following
an approach similar to that described in \cite{2012ApJ...749..156O}
by including an effective adiabatic index\footnote{This
simplified approach takes into account the macroscopic effect caused
by particle acceleration, but does not exhaust all possible effects
of the fast particles.} $\gamma\rs{eff}$ which depends on the
injection rate of particles $\eta$ (i.e., the fraction of CSM
particles with momentum above a threshold value, $p\rs{inj}$, that
are involved in the acceleration process; \citealt{2005MNRAS.361..907B})
but neglecting nonlinear magnetic-field amplification.

The SNR simulations were performed using the \PLUTO\ code
(\citealt{2007ApJS..170..228M, 2012ApJS..198....7M}) configured to
compute intercell fluxes with a two-shock Riemann solver (the
linearized Roe Riemann solver in the case of HD simulations and the
HLLD approximate Riemann solver in the case of MHD simulations; see
Paper I for more details). The HD/MHD equations were solved in a
3D Cartesian coordinate system $(x,y,z)$, assuming the Earth vantage
point to lie on the negative $y$-axis. The remnant is oriented in
such a way that the Fe-rich fingers developed soon after the core
bounce point toward the same direction as the extended Fe-rich
regions observed in \casa\ (see Paper I). The large physical scales
spanned from the shock breakout to the full-fledged remnant at the
age of 2000~years were followed by gradually extending the computational
domain (a Cartesian box covered by a uniform grid of $1024^3$ zones)
as the forward shock propagates outward. The spatial resolution
varies between $\approx 2.3 \times 10^{11}$~cm (on a domain extending
between $-1.2 \times 10^{14}$~cm and $1.2 \times 10^{14}$~cm in all
directions) at the beginning of the calculation to $\approx 0.018$~pc
(on a domain extending between $-9.4$~pc and $9.4$~pc) at the end.

\begin{table}
\caption{Setup for the simulated models that best match the observations$^a$.}
\label{tabmod}
\begin{center}
\begin{tabular}{llllll}
\hline
\hline
 & \multicolumn{2}{l}{Parameter}  &  \multicolumn{3}{l}{Value}  \\
\hline
SN model & \multicolumn{2}{l}{$E\rs{exp}$} &  \multicolumn{3}{l}{$1.5\times 10^{51}$ erg $=1.5$~B}  \\
~~~~W15-2-cw-IIb$^{b}$  & \multicolumn{2}{l}{$M\rs{ej}$}  &  \multicolumn{3}{l}{$3.3\,M_{\odot}$}  \\
    & \multicolumn{2}{l}{$E\rs{exp}/M\rs{ej}$} &  \multicolumn{3}{l}{$0.45$~B$/M_{\odot}$}  \\
\\
SNR Model  &  rad.      &  {\bf B} &  $\eta$  &   $n\rs{w}$  & shell\\
           &  dec.      &       &          &  cm$^{-3}$ \\ \hline
W15-2-cw-IIb-HD$^{c}$	&  no	&  no	&  0	     &  0.8  & no  \\
W15-IIb-sh-HD           &  no   &  no   &  0         &  0.8  & SH1$^{d}$ \\
W15-IIb-sh-HD-1eta      &  no   &  no   &  $10^{-4}$ &  0.7  & SH1 \\
W15-IIb-sh-HD-10eta     &  no   &  no   &  $10^{-3}$ &  0.6  & SH1 \\
W15-IIb-sh-HD+dec       &  yes  &  no   &  0         &  0.8  & SH1 \\
W15-IIb-sh-MHD+dec      &  yes  &  yes  &  0         &  0.8  & SH1 \\
W15-IIb-sh-HD-1eta-sw	&  no	&  no	&  $10^{-4}$ &  0.7  & SH2$^{e}$ \\
W15-IIb-sh-HD-1eta-az	&  no	&  no	&  $10^{-4}$ &  0.7  & SH3$^{f}$ \\
\hline
\end{tabular}
\end{center}
$^{a}$ We ran about fifty 3D high-resolution simulations of the
SNR, exploring the space of parameters reported in Table~\ref{tabshell};
we summarize here only the models that best match the observations.
$^{b}$ Presented in \cite{2017ApJ...842...13W}.
$^{c}$ Presented in Paper I.
$^{d}$ The shell is characterized by the best-fit parameters reported
in Table~\ref{tabshell} but with $n\rs{sh} = 10$~cm$^{-3}$ and $\phi = 0$.
$^{e}$ The shell is the same as in SH1 but rotated by $90^{\rm o}$
clockwise about the $y$ axis.
$^{f}$ The shell is the same as in SH1 but with density $n\rs{sh}
= 20$~cm$^{-3}$ and $\phi = 50^{\rm o}$ (see Table~\ref{tabshell}).
\end{table}

First we evaluated the effects of back-reaction of accelerated
cosmic rays on the results by considering simulations either with
(models W15-IIb-sh-HD-1eta and W15-IIb-sh-HD-10eta) or without
(W15-IIb-sh-HD) the modifications of the shock dynamics due to
cosmic rays acceleration at both the forward and reverse shocks.
We considered two cases: $\eta = 10^{-4}$ and $\eta = 10^{-3}$,
leading to $\gamma\rs{eff} \approx 3/2$ (model W15-IIb-sh-HD-1eta)
and $\gamma\rs{eff} \approx 4/3$ (model W15-IIb-sh-HD-10eta),
respectively (see Fig.~2 in \citealt{2016ApJ...822...22O}). The
former case is the most likely for \casa, according to
\cite{2016ApJ...822...22O}; the latter is an extreme case of a very
efficient particle acceleration. For the sake of simplicity, in the
present calculations we did not assume a time dependence of
$\gamma\rs{eff}$, i.e., we assumed that the lowest value is reached
immediately at the beginning of the simulation.

In Paper I, we found that the energy deposition from radioactive
decay provides an additional pressure to the plasma which inflates
ejecta structures rich in decaying elements. Thus, we performed an
additional simulation (W15-IIb-sh-HD+dec) analogous to model
W15-IIb-sh-HD but including the effects of energy deposition from
radioactive decay. We investigated these effects on the remnant-shell
interaction by comparing models W15-IIb-sh-HD and W15-IIb-sh-HD+dec.

We have also investigated the possible effects of an ambient magnetic
field on the remnant-shell interaction. In fact, although the
magnetic field does not affect the overall evolution of the remnant,
it may limit the growth of HD instabilities that develop at the
contact discontinuity or during the interaction of the forward shock
with inhomogeneities of the CSM (see, for instance,
\citealt{2019A&A...622A..73O}) as, in the present case, the
circumstellar shell. Hence, following Paper I, we performed a
simulation as model W15-IIb-sh-HD+dec but including an ambient
magnetic field (model W15-IIb-sh-MHD+dec) and compared the two
models. As in Paper I, we adopted the ambient magnetic field
configuration described by the ``Parker spiral'' resulting from the
rotation of the progenitor star and from the corresponding expanding
stellar wind (\citealt{1958ApJ...128..664P}); the adopted pre-SN
magnetic field has an average strength at the stellar surface
$B\rs{0} \approx 500$~G (\citealt{2009ARA&A..47..333D}). 

A few words of caution are needed about the adopted magnetic field.
In fact, the pre-SN field strength and configuration are unknown
in \casa\ and our choice is, therefore, arbitrary.  Furthermore,
typical magnetic field strengths in post-shock plasma inferred from
observations of \casa\ are of the order of $0.5$~mG (e.g.,
\citealt{2018ApJ...853...46S}), whereas the pre-SN magnetic
field is of the order of $0.2~\mu$G at a distance of 2.5~pc from
the center of explosion and the highest values of magnetic field
strength in post-shock plasma at the age of \casa\ are of the order
of $10~\mu$G in our MHD simulation (model W15-IIb-sh-MHD+dec).
These observations suggest that some mechanism of magnetic field
amplification is at work, such as turbulent motion in the post-shock
plasma (e.g., \citealt{2007ApJ...663L..41G, 2009ApJ...695..825I}
and references therein) and/or non-linear coupling between cosmic
rays and background magnetic field (\citealt{2004MNRAS.353..550B}).
The former requires quite high spatial resolution and the latter
solving the evolution of the non-linear coupling in a short time-scale
and length-scale. In fact, both these mechanisms are not included
in our MHD simulations that describe the evolution of the whole
remnant. In light of this, we expect that stronger fields may
have effects on the remnant-shell interaction not included in our
models especially for the acceleration of particles (most likely
correlated with the magnetic field amplification) and the growth
of HD instabilities. Consequently, the synthesis of radio emission
presented in Sect.~\ref{rs_asym} is expected to predict radio images
which cannot be directly compared with radio observations of \casa.
Nevertheless, these synthetic maps were derived with the aim to
identify the position of the reverse shock during the remnant-shell
interaction. Since the position and resulting overall shape of the
reverse shock do not depend on the particular configuration of the
magnetic field adopted, the maps can be safely used for our purposes.

\subsection{The inhomogeneous CSM}
\label{sec:csm}

In Paper I, the remnant was described as expanding through the
spherically symmetric wind of the progenitor star. The wind density
was assumed to be proportional to $r^{-2}$ (where $r$ is the radial
distance from the center of explosion) and was equal to $n\rs{w} =
0.8$~cm$^{-3}$ (consistent with the values of post-shock wind
density inferred from observations of \casa;
\citealt{2014ApJ...789....7L}) at the radius $r\rs{fs} = 2.5$~pc,
namely the nominal current outer radius of the remnant (at a distance
of $\approx 3.4$~kpc). Assuming a wind speed of $10-20$~km~s$^{-1}$
(typical values for the wind during the red supergiant phase), the
estimated mass-loss rate is $\dot M \approx 2-4 \times
10^{-5}\,M_{\odot}$~yr$^{-1}$. Furthermore, a progressive flattening
of the wind profile to a uniform density $n\rs{c}= 0.1$~cm$^{-3}$
was considered at distances $> 3$~pc (where we ignore the
structure of the still unshocked CSM) to prevent unrealistic low
values of the density.

Here, we aim at exploring the effects of a dense shell of CSM on
the evolution of the remnant and at testing the hypothesis
that some of the features observed in the reverse shock of \casa\
may be interpreted as signatures of a past interaction of the remnant
with an asymmetric circumstellar shell. Deriving an accurate
reconstruction of the pre-SN CSM around \casa\ is well beyond the
scope of the paper. Thus, for our purposes, we adopted an
idealized and parametrized description of the CSM which consists
of a spherically symmetric wind (as in Paper I) and a shell of
material denser than the wind. We also allowed the shell to be
asymmetric with one hemisphere being denser than the other. In fact,
several lines of evidence suggest that the CSM around massive stars
can be characterized by the presence of asymmetric and dense
circumstellar shells, resulting from episodic massive eruptions
during the late stages of star evolution (e.g.,
\citealt{2014MNRAS.438.1191S, 2014AJ....147...23L, 2014ApJ...787..163G}).

The asymmetry was modeled with an exponential density stratification
along a direction with unit vector $\vec D$, which defines the
symmetry axis of the shell. This dipole asymmetry with the
enhancement covering 2$\pi$ steradians is made simply to differentiate
the two hemispheres of the shell; one might expect an enhancement
occupying a smaller solid angle (as seen from the explosion center)
to produce similar effects on the reverse shock dynamics over a
smaller range of azimuth. The transition from the shell to
the wind was modulated by a Gaussian function. The exponential and
Gaussian functions were selected to allow for a smooth transition
between the wind and the shell. Considering the orientation of the
remnant in the 3D Cartesian coordinate system (see Sect.~\ref{sec:sn-snr}),
the density distribution of the CSM is given by:

\begin{equation}
n = n\rs{w} \left(\frac{r\rs{fs}}{r}\right)^2 + 
n\rs{sh}\exp\left[-\frac{(r-r\rs{sh})^2}{2\sigma^2}\right]
\exp\left[\frac{\vec r \cdot \vec D}{H}\right]\,,
\label{eq:csm}
\end{equation}

\noindent
where $n\rs{w}$ has a value between $0.6$~cm$^{-3}$ and $0.8$~cm$^{-3}$,
depending on the injection efficiency (see \citealt{2016ApJ...822...22O}),
$r\rs{fs} = 2.5$~pc, $n\rs{sh}$ is a reference density of the shell,
$r\rs{sh}$ is the shell radius, $\sigma$ represents the shell
thickness, $H$ is the scale length of the shell density, $\vec
r \cdot \vec D = x \cos \theta \cos \phi - y \sin \phi + z \sin
\theta \cos \phi$, and $\theta$ and $\phi$ are the angles measured:
the former in the $[x, z]$ plane (i.e., around the $y$-axis)
counterclockwise from the $x$-axis (i.e., from the west) and the
latter about the $z$-axis counterclockwise from the $[x, z]$
plane\footnote{So that the unit vector $\vec D$ has components:
$D\rs{x} = \cos \theta \cos \phi$, $D\rs{y} = -\sin \phi$ and
$D\rs{z} = \sin \theta \cos \phi$.} (i.e., from the plane of the
sky). We note that the parameter $H$ determines the contrast
between the densest and least dense portions of the shell and,
therefore, the level of asymmetry introduced between the two remnant
hemispheres.

\begin{figure}[!t]
  \begin{center}
    \leavevmode
        \epsfig{file=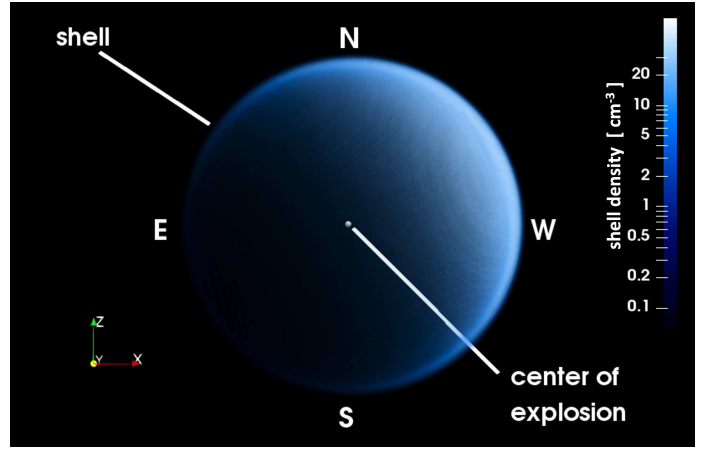, width=9cm}
	\caption{Schematic representation of the massive
	circumstellar shell. The center of the explosion is coincident
	with the geometric center of the spherical shell. The Earth
	vantage point lies on the negative $y$-axis. In this
	representation the angles in Eq.~\ref{eq:csm} are
	$\theta = 30^{\rm o}$ and $\phi = 0$ and the shell has the
	maximum density in the NW quadrant and the minimum density
	in the SE quadrant.}
  \label{fig1}
\end{center} \end{figure}

As a first step, we explored the space of parameters of the
shell, assuming that $\vec D$ (i.e., the symmetry axis of the shell)
lies in the plane of the sky ($[x, z]$ plane) and, therefore, is
perpendicular to the line-of-sight (LoS). In this case, $\phi = 0$
and $\vec r \cdot \vec D = x \cos \theta + z \sin \theta$. An example
of pre-SN CSM resulting in this last case, for $\theta = 30^{\rm o}$,
is shown in Fig.~\ref{fig1}. Considering that the Earth vantage
point lies on the negative $y$-axis, the shell has the maximum
density in the north-west (NW) quadrant and the minimum density in
the south-east (SE) quadrant.

As a second step, we explored the possibility that $\vec D$ forms
an angle $\phi > 0$ with the plane of the sky in Eq.~\ref{eq:csm}.
In other words, we explored models in which the shell was also
denser in its blueshifted nearside than in the redshifted farside,
as suggested by the evidence that the large majority of QSFs is at
blue shifted velocities (\citealt{1995ApJ...440..706R}). In
Sect.~\ref{dop_prj}, we discuss the results of this exploration and
present model W15-IIb-sh-HD-1eta-az, our favorite model (with
$\phi = 50^{\rm o}$) to describe the dynamics of the reverse and
forward shocks observed in \casa.

\section{Results}
\label{sec:results}

We ran 3D high-resolution simulations of the SNR, searching for the
parameters of the shell (density, radius, and thickness) and of its
degree of asymmetry (the angles $\theta$ and $\phi$ and the density
scale length) which can reproduce the slow down of the reverse shock
velocity in the western region of \casa, the offset between the
geometric centers of the reverse and forward shocks as inferred
from the observations, and the evidence that the nonthermal emission
from the reverse shock in the western region is brighter than in
the eastern region, simultaneously.  Table~\ref{tabshell} reports
the shell parameters of our simulations most closely reproducing
the observations and the range of values explored. The Table
also reports the total mass calculated for the shell. We note that
this mass depends on the geometry of the shell adopted. We expect
that a smaller or partial shell or a shell with a shape deviating
from spherical symmetry (for instance, more elongated on one side)
may produce similar observables with a different mass.

In Sect.~\ref{interaction}, we discuss in detail the interaction
of the remnant with the dense shell for one of our simulations with
$\phi = 0$ in Eq.~\ref{eq:csm} more closely resembling the observations
(model W15-IIb-sh-HD-1eta). Then, in Sect.~\ref{param_space} we
describe how the remnant evolution changes using different shell
parameters. Finally, in Sects.~\ref{rs_asym} and \ref{dop_prj} we
analyze the remnant asymmetries caused by the interaction of the
remnant with the shell, including also the case with $\phi > 0$ (in
Sect.~\ref{dop_prj}), and compare the model results with observations.

\begin{table}
\caption{Explored and favorite parameters of the shell in the CSM for
the models of \casa$^{a}$. The favorite parameters correspond to
those of model W15-IIb-sh-HD-1eta-az in Table~\ref{tabmod}.}
\label{tabshell}
\begin{center}
\begin{tabular}{llll}
\hline
\hline
Parameter  &  Units  & Range     & Favorite \\
           &         & Explored  & Value \\
\hline
$n\rs{sh}$  &  (cm$^{-3}$)   &   $[5; 20]$                     & $20^{b}$   \\
$r\rs{sh}$  &  (pc)          &   $[1; 2]$                      & $1.5$  \\
$\sigma$    &  (pc)          &   $[0.02; 0.05]$                & $0.02$ \\
$\theta$    &                &   $[35^{\rm o}; -57^{\rm o}]$   & $30^{\rm o}$  \\
$\phi$      &                &   $[0^{\rm o}; 50^{\rm o}]$     & $50^{\rm o}$  \\
$H$         &  (pc)          &   $[0.35; 3.5]$                 & $0.7$   \\ \hline
$M\rs{sh}$  &  ($M_{\odot}$) &   $[0.43; 3.71]$                & $2.3^c$ \\ 
$M\rs{SE}$  &  ($M_{\odot}$) &   $[0.16; 0.49]$                & $0.7$ \\ 
$M\rs{NW}$  &  ($M_{\odot}$) &   $[0.22; 3.22]$                & $1.6$ \\ 
$n\rs{SE}$  &  (cm$^{-3}$)   &   $[1.6; 5.8]$                  & $6.3$  \\
$n\rs{NW}$  &  (cm$^{-3}$)   &   $[6.7; 89]$                   & $77$   \\
\hline
\end{tabular}
\end{center}
$^{a}$ The reference density of the shell, $n\rs{sh}$, the shell
radius and thickness, $r\rs{sh}$ and $\sigma$, the angle $\theta$
(measured in the plane of the sky, about the $y$-axis,
counterclockwise from the west), the angle $\phi$ (measured about
the $z$-axis counterclockwise from the plane of the sky) and the
scale length of the shell density, $H$, are used in Eq.~\ref{eq:csm}
to describe the CSM. The total mass of the shell, $M\rs{sh}$, the
masses of the SE and NW hemispheres of the shell, $M\rs{SE}$ and
$M\rs{NW}$, and the peak densities of the shell in the SE and NW
directions, $n\rs{SE}$ and $n\rs{NW}$, are derived from the
simulations.
$^{b}$ The density $n\rs{sh}=10$~cm$^{-3}$ if $\phi = 0$.
$^{c}$ The mass of the shell is $M\rs{sh} = 1.3\,M_{\odot}$ if $\phi
= 0$.
\end{table}

\subsection{Interaction of the remnant with the dense shell}
\label{interaction}

\begin{figure*}[!ht]
  \begin{center}
    \leavevmode
        \epsfig{file=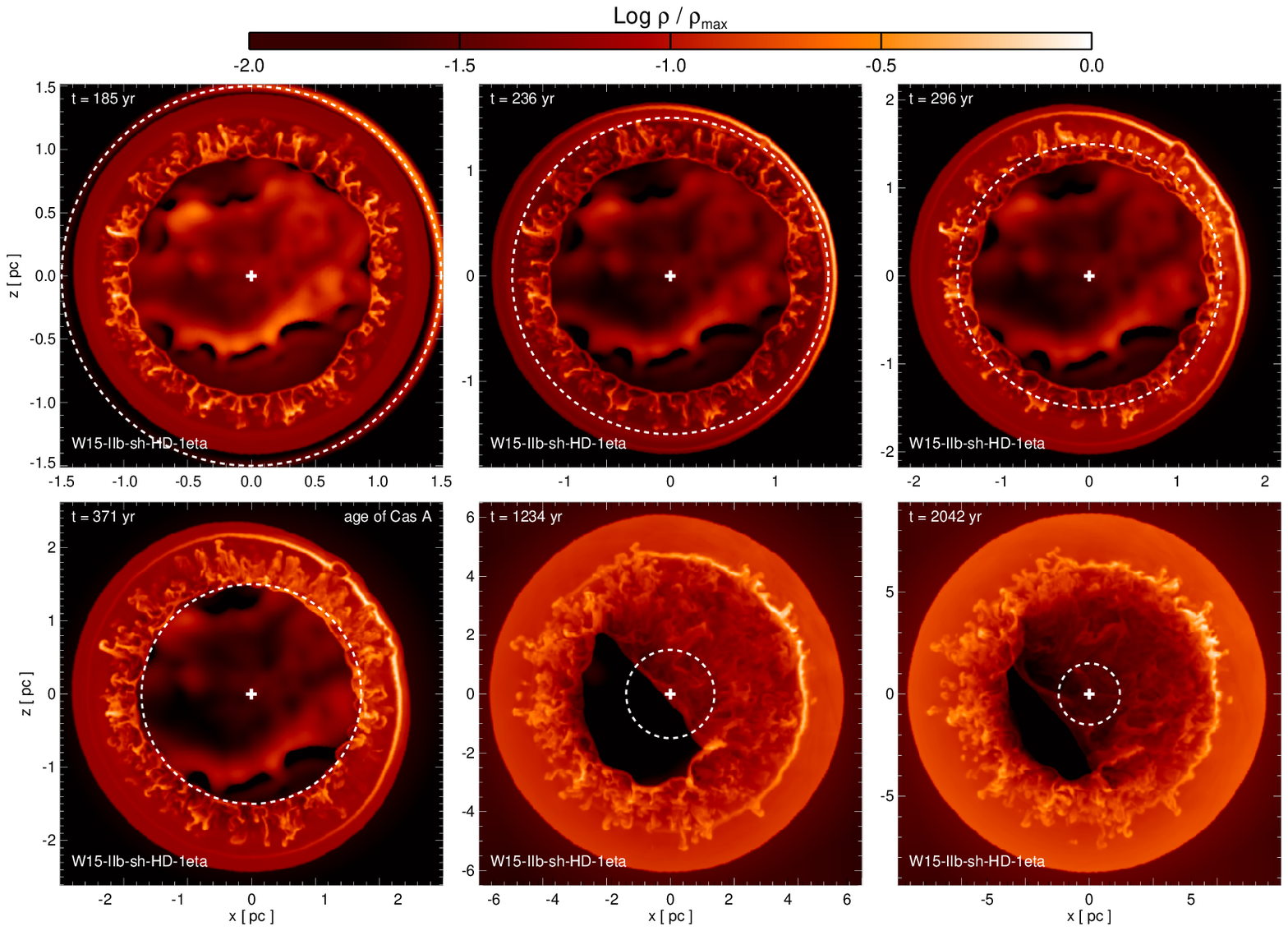, width=18.cm} 
	\caption{2D sections in the $(x, z)$ plane of the spatial
	density distribution of shocked plasma (in log scale) at
	the labeled times for model W15-IIb-sh-HD-1eta. Each image
	has been normalized to its maximum, $\rho\rs{max}$, for
	visibility. The dashed circle in each panel marks the
	pre-shock position of the dense shell in the $(x, z)$ plane;
	the cross shows the center of the explosion.}
  \label{map_dens}
\end{center} \end{figure*}

The evolution of the remnant in the phase before the interaction
of the blast wave with the shell is analogous to that presented in
Paper I. Initially the metal-rich ejecta expand almost homologously,
though in models including the radioactive decay significant
deviations are present in the innermost ejecta (rich in $^{56}$Ni
and $^{56}$Co) due to heating caused by the decay chain $^{56}{\rm
Ni} \rightarrow\ ^{56}{\rm Co} \rightarrow\ ^{56}{\rm Fe}$. These
effects decrease fast with time and are significant only during the
first year of evolution. Thus, after this initial inflation of
Fe-rich ejecta, we expect that the qualitative evolution of the
remnant is similar in models either with or without the radioactive
decay effects included (see Paper I). In all the models, the
Fe-group elements start to interact with the reverse shock about
$\approx 30$ years after the SN when almost all $^{56}$Ni and
$^{56}$Co have already decayed to stable $^{56}$Fe (see Paper I for
details).

\begin{figure}[!t]
  \begin{center}
    \leavevmode
        \epsfig{file=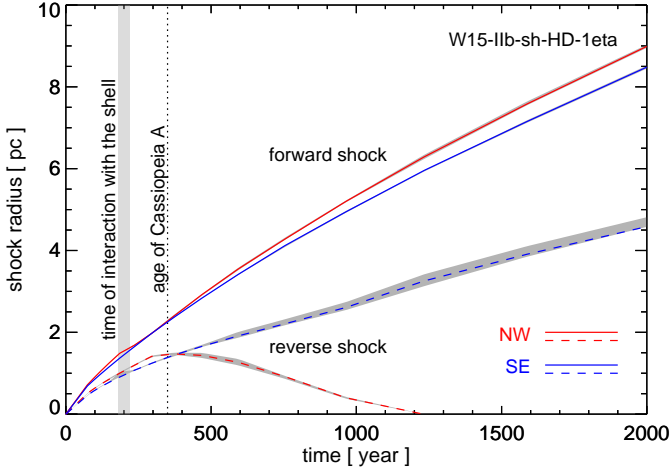, width=9cm}
	\caption{Radii of the forward (solid lines) and reverse
	(dashed lines) shocks versus time in the NW (red lines) and
	SE (blue lines) hemispheres for model W15-IIb-sh-HD-1eta.
	The shock positions are averaged over solid angles
	(as seen from the explosion center) of $10^{\rm o}$ around
	the NW and SE directions. The vertical dotted line marks
	the approximate age of \casa; the shaded rectangular
	box marks the time of remnant-shell interaction. The
	grey bands under the colored average curves indicate the
	range of shock positions at each time in the solid angles
	examined.}
  \label{shock_pos}
\end{center} \end{figure}

In our simulations best matching the observations (all reported in
Table~\ref{tabmod}), the dense shell of CSM has a radius $r\rs{sh}
= 1.5$~pc. In model W15-IIb-sh-HD-1eta, the blast wave starts to
interact with the shell $\approx 180$~years after the SN event (see
upper left panel in Fig.~\ref{map_dens}). The NW side of the shell
is hit first due to the large-scale asymmetries of the blast wave
inherited from the earliest moments of the explosion\footnote{Note
that the large-scale asymmetry of the SN explosion points to the
NW in the redshifted side of the remnant (see Paper I), while the
densest part of the shell lies in the NW quadrant of the plane of
the sky. Their projections into the plane of the sky coincide by
chance.} (see Paper I). At the beginning of the interaction, the
forward shock slows down because of the propagation through a medium
denser than the $r^{-2}$ wind density distribution. Consequently,
the distance between the forward and reverse shock gradually
decreases. This effect is the largest in the NW side which is hit
earliest and where the CSM shell in our reference model has the
highest density. This is also evident in Fig.~\ref{shock_pos} showing
the forward and reverse shock radii in the NW and SE hemispheres
of the remnant: the forward shock in the NW side shows a slow down
in its expansion at the time of interaction with the shell that is
not present in the SE side. This further enhances the degree of
asymmetry in the remnant, leading to a thickness of the mixing
region between the forward and reverse shocks that is the smallest
in the NW side. The shell is fully shocked at $t \approx 240$~years
(upper center panel in Fig.~\ref{map_dens}). At later times, the
forward shock travels again through the $r^{-2}$ wind density
distribution and its velocity gradually increases to the values
expected without the interaction with the shell.

\begin{figure}[!th]
  \begin{center}
    \leavevmode
        \epsfig{file=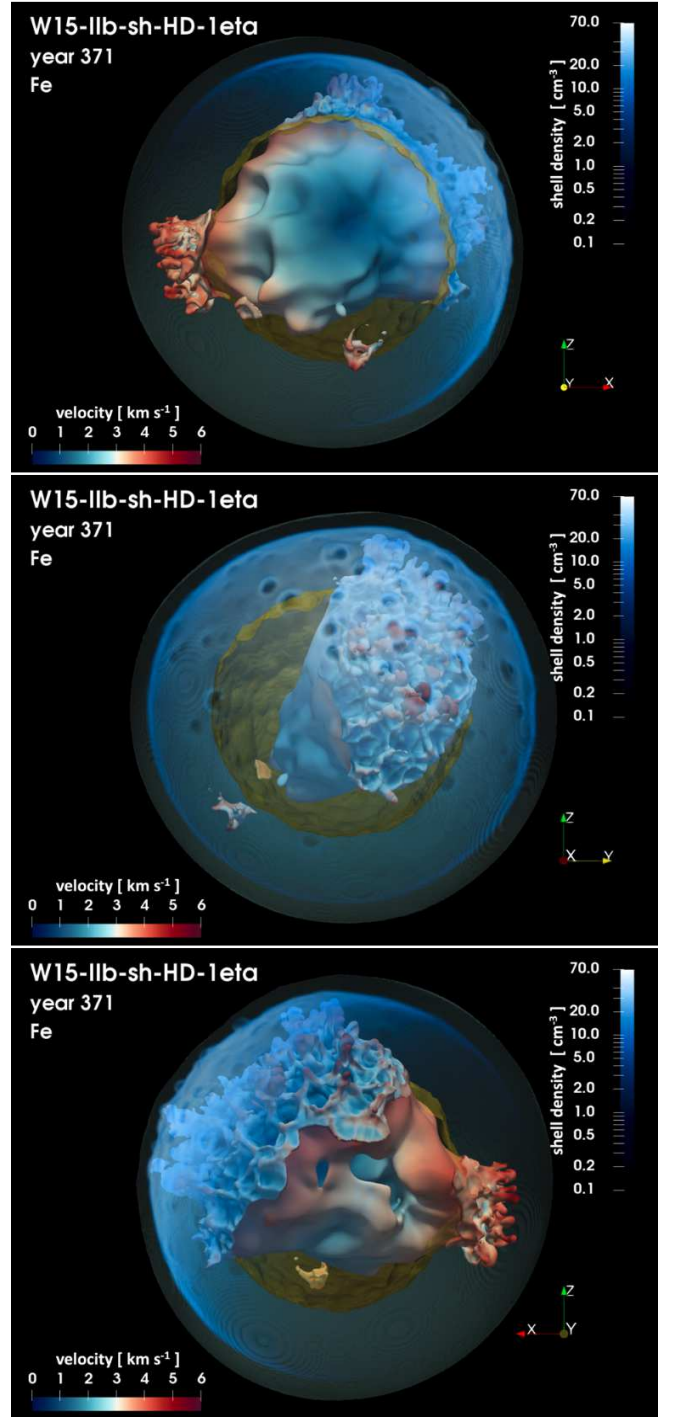, width=8.5cm}
	\caption{Isosurface of the distribution of Fe (corresponding
	to a value of Fe density which is at 5\% of the peak density)
	at the age of \casa\ for different viewing angles for model
	W15-IIb-sh-HD-1eta; the colors give the radial velocity in
	units of 1000~km~s$^{-1}$ on the isosurface (color coding
	defined at the bottom of each panel). The semi-transparent
	clipped quasi-spherical surfaces indicate the forward (green)
	and reverse (yellow) shocks. The shocked shell is visualized
	through a volume rendering that uses the blue color palette
	(color coding on the right of each panel); the opacity is
	proportional to the plasma density. A navigable 3D graphic
	of this model is available at https://skfb.ly/o8FnO.}
  \label{fig_evol}
\end{center} \end{figure}

The remnant-shell interaction drives a reflected shock that travels
inward through the mixing region and that is the most energetic
where the shell is the densest. The inward propagating shock wave
reaches the reverse shock at $t \approx 290$~years (upper right
panel in Fig.~\ref{map_dens}). As a result, the reverse shock
velocity in the observer frame decreases and, again, the effect is
the largest (with velocities equal to zero or even negative) in the
NW side where the reflected shock is the most energetic
(Fig.~\ref{shock_pos}). At the age of \casa, the reverse shock in
the NW hemisphere has already started to move inward in the observer
frame (see Fig.~\ref{shock_pos}) and the remnant shows the effects
of the interaction with the shell (lower left panel in
Fig.~\ref{map_dens}). As a result, the mixing region is less
extended and the density of the shocked plasma is higher in the
western than in the eastern region.

Fig.~\ref{fig_evol} shows the spatial distribution of Fe at the age
of \casa\ in model W15-IIb-sh-HD-1eta. The effects of back-reaction
of accelerated cosmic rays do not change significantly this
distribution (models W15-IIb-sh-HD and W15-IIb-sh-HD-10eta show
similar results), whilst the decay of radioactive species leads to
the inflation of the Fe-rich plumes in models W15-IIb-sh-HD+dec and
W15-IIb-sh-MHD+dec (see Paper I). The figure shows different viewing
angles, namely with the perspective on the negative $y$-axis (i.e.,
the vantage point is at Earth; upper panel), on the positive $x$-axis
(middle panel), and on the positive $y$-axis (i.e., the vantage
point is from behind \casa; lower panel). At this time about 35\%
of Fe has already passed through the reverse shock (see Fig.~5 in
Paper I), leading to the formation of large regions of shocked
Fe-rich ejecta in coincidence with the original large-scale fingers
of Fe-group elements (see Paper I for more details). The shocked
dense shell of CSM has already started to interact with the fingers
of ejecta developed by HD instabilities (Rayleigh-Taylor,
Richtmyer-Meshkov, and Kelvin-Helmholtz shear instability;
\citealt{1973MNRAS.161...47G, 1991ApJ...367..619F, 1992ApJ...392..118C}).
In particular, the figure shows Fe-rich shocked filamentary structures,
which extend from the contact discontinuity toward the forward
shock. These fingers protrude into the shocked shell material,
producing holes in the shell (see middle panel in Fig.~\ref{fig_evol})
and driving the mixing between stellar and shell material. The
fingers in the NW side are closer to the forward shock than in the
SE side, due to the reduced distance between the contact discontinuity
and the forward shock where the shell is the densest. The radial
velocity of the fingers in the NW side is smaller than in the SE
(see the color code of the isosurface in Fig.~\ref{fig_evol}) due
to the passage of the inward shock, which is the most energetic in
the NW. This is consistent with observations (e.g.,
\citealt{2002A&A...381.1039W}).

At later times, the degree of asymmetry of the reverse shock structure
largely increases due to the fastest inward propagation of the
reverse shock in the NW region (see Fig.~\ref{shock_pos}). At the
age of $\approx 1000$~yr, the reverse shock is highly asymmetric
and reaches the center of the explosion from NW (see lower center
panel in Fig.~\ref{map_dens} and Fig.~\ref{shock_pos}). Then, it
starts to propagate through the ejecta traveling outward in the SE
portion of the remnant (lower right panel in Fig.~\ref{map_dens}).
In the meantime, the forward shock continues to travel through the
$r^{-2}$ wind density distribution with roughly the same velocity
in all directions. The forward shock appears to be spherically
symmetric at the end of the simulation, when the remnant has a
radius $R \approx 9$~pc and an age of $\approx 2000$~yr. It is
interesting to note that, at this age, the signatures of the
remnant-shell interaction are clearly visible in the reverse shock,
whilst the forward shock apparently does not keep memory of the
past interaction (see Fig.~\ref{shock_pos}).

\begin{figure*}[!t]
  \begin{center}
    \leavevmode
        \epsfig{file=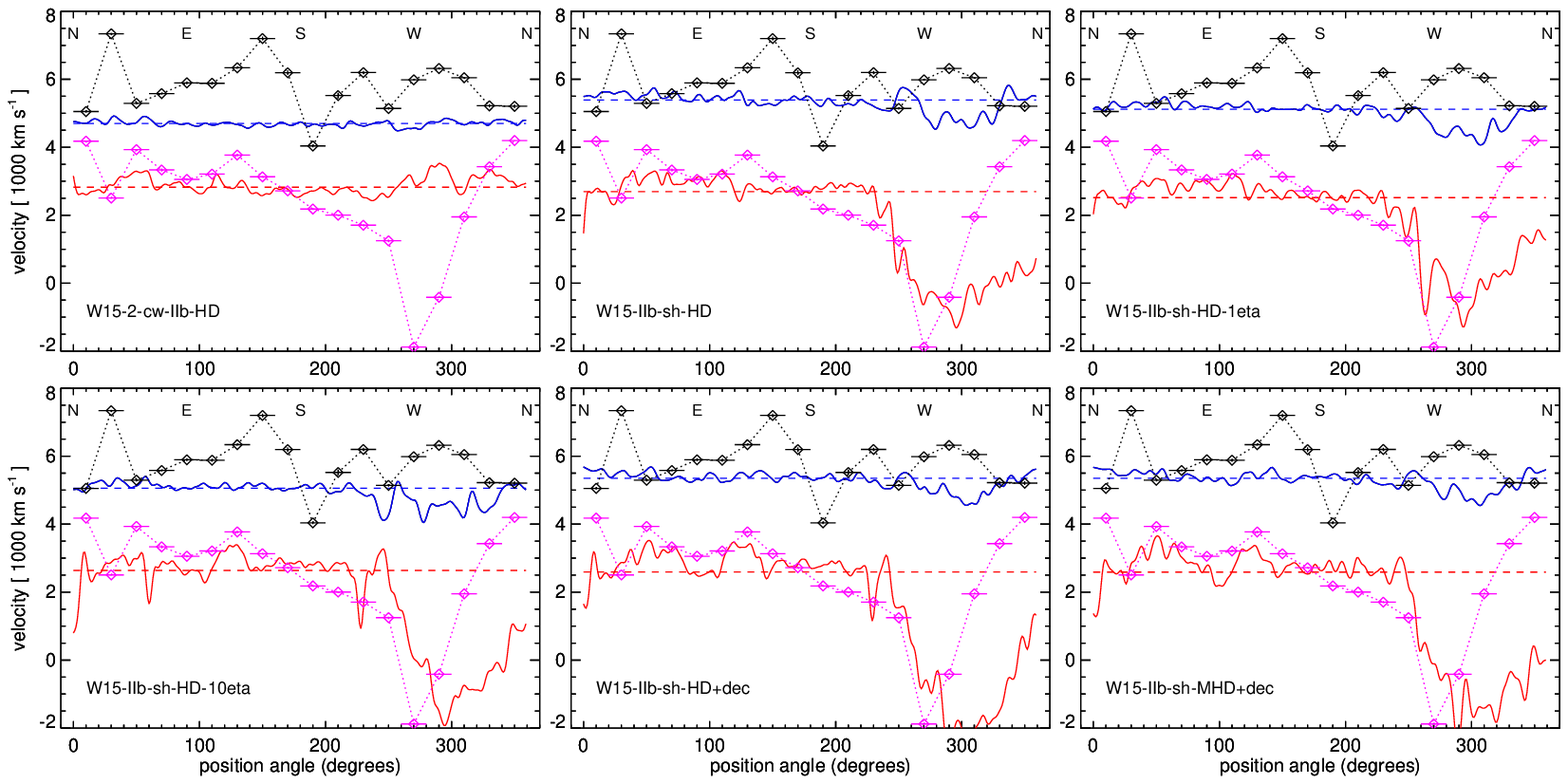, width=18cm}
	\caption{Forward (blue) and reverse (red) shock velocities
	versus the position angle in the plane of the sky at the
	age of \casa\ for the first six SNR models listed in
	Table~\ref{tabmod}, including model W15-2-cw-IIb-HD presented
	in Paper I. The dashed horizontal lines mark the median
	values of the respective velocities. The velocities of the
	forward (black diamonds) and reverse (magenta diamonds)
	shocks derived from the analysis of Chandra observations
	(\citealt{2022ApJ...929...57V}) are overplotted for
	comparison.}
  \label{prof_vsh}
\end{center} \end{figure*}

\subsection{Effects of shell parameters on the remnant evolution}
\label{param_space}

We explored the space of parameters of the shell by performing about
fifty simulations. The parameters explored are (see Table~\ref{tabshell}):
the reference density of the shell, $n\rs{sh}$, the shell radius
and thickness, $r\rs{sh}$ and $\sigma$, the angle $\theta$ (measured
counterclockwise from the west; see Eq.~\ref{eq:csm}), and the scale
length of the shell density, $H$. For this exploration, the angle
$\phi$ was fixed equal to zero; its effect is investigated in
Sect.~\ref{dop_prj}. The parameter space was explored adopting an
iterative process of trial and error to converge on model parameters
that qualitatively reproduce the profiles of the forward and reverse
shock velocities versus the position angle in the plane of the sky
at the age of \casa\ (e.g., \citealt{2022ApJ...929...57V}; see also
Fig.~\ref{prof_vsh} and Sect.~\ref{rs_asym}). We note that our
models do not pretend to be able to reconstruct the structure of
the pre-SN circumstellar shell but they aim to test the
possibility that the inward-moving reverse shock observed in the
western hemisphere of \casa\ can be interpreted as the signature
of a past interaction with a circumstellar shell. From our exploration,
we found that the models producing velocity profiles, which more
closely reproduce the observations (listed in Table~\ref{tabmod}),
are characterized by the common set of parameters listed in
Table~\ref{tabshell} with $n\rs{sh}=10$~cm$^{-3}$ if $\phi = 0$
(see the favorite values).

The exploration of the parameter space was limited to asymmetric
shells with the densest side in the western hemisphere of the
remnant, namely where the profile of the reverse shock velocity has
a minimum (see Fig.~\ref{prof_vsh}). The shell asymmetry is regulated
by the angle $\theta$ and the scale length of the shell density,
$H$. The former parameter determines where the shell is the densest
in the plane of the sky, so that the effects of interaction with
the shell are the largest and the reverse shock velocity has a
minimum. Larger (lower) values of $\theta$ determine a shift of the
minimum shock velocity toward the north (south) in Fig.~\ref{prof_vsh}.
The parameter $H$ regulates the density contrast between the
two hemispheres of the shell (namely its densest and least dense
portions) and, therefore, the level of asymmetry introduced by the
shell: the higher the value of $H$ the smaller the contrast between
the reverse shock velocities in the two remnant hemispheres.

In simulations assuming a shell radius smaller than $r\rs{sh}
= 1.5$~pc, the remnant starts to interact with the shell at an
earlier time. Hence, the reverse shock starts earlier to move inward
where the shell is dense, leading to a reverse shock structure
that significantly deviates from the spherical shape at the age of
\casa\ (this happens at later times in our favorite simulations
listed in Table~\ref{tabmod}): a result which is at odds with
observations. On the other hand, in simulations with a higher shell
radius, the forward shock in the western hemisphere still has
expansion velocities much lower than those in the eastern hemisphere
(producing an evident minimum in the profiles in Fig.~\ref{prof_vsh})
because it did not have the time to re-accelerate to the velocity
values expected when it propagates through the wind of the progenitor
star.

The shell density regulates the slow down of the forward shock
traveling through the shell and the strength of the inward shock.
In models with a shell denser than in our favorite cases (listed
in Table~\ref{tabmod}), the slow down of the forward shock is higher,
leading to a minimum in its velocity profile at the age of \casa\
(at odds with observations; see Fig.~\ref{prof_vsh}), and the inward
shock is stronger, leading to a deeper minimum in the profile
of the reverse shock (inward velocities lower than the minimum
values observed, namely $\approx -2000$~km~s$^{-1}$). On the other
hand, models with a shell less dense than in our favorite models
produce a minimum in the velocity profile of the reverse shock with
velocities higher than observed ($> -2000$~km~s$^{-1}$). A similar
role is played by the thickness of the shell, $\sigma$: a thicker
(thinner) shell produces larger (smaller) effects on the forward
and reverse shock dynamics. In principle, one could trade off density
and thickness to produce similar results. However, we found that,
in simulations with higher values of $\sigma$, the reverse shock
deviates from the spherical shape at the age of \casa\ if the
interaction of the remnant with the thicker shell starts at earlier
times than in our favorite models; conversely, if the interaction
starts at the time of our favorite simulations, the forward shock
shows a minimum in its velocity profile because it left the thicker
shell too late and it did not have time to re-accelerate.

Including the effects of radioactive decay does not qualitatively
change the evolution of the remnant and its interaction with the
shell. As shown in Paper I, the energy deposition by radioactive
decay provides additional pressure to the plasma, which inflates
structures with a high mass fraction of decaying elements against
the surroundings. Thus, the expansion of the ejecta is powered by
this additional pressure and the remnant expands slightly faster
than in the case without these effects taken into account. As a
consequence, the remnant starts to interact with the shell slightly
earlier in models W15-IIb-sh-HD+dec and W15-IIb-sh-MHD+dec than
in the others. The reflected shock from the shell reaches the reverse
shock at earlier times. Thus, at the age of \casa, the reverse shock
in the NW region moves inward in the observer frame with slightly
higher velocities than in models without radioactive decay (e.g.,
compare models W15-IIb-sh-HD and W15-IIb-sh-HD+dec in
Fig.~\ref{prof_vsh}). The differences between the models, however,
are moderate ($< 30$\%).

As for the effect of an ambient magnetic field, it does not, as
expected, influence the overall dynamics of the forward and reverse
shocks. Model W15-IIb-sh-MHD+dec (the only one including the magnetic
field) shows an evolution similar to that of model W15-IIb-sh-HD+dec.
The main effect of the magnetic field is to limit the growth of HD
instabilities at the contact discontinuity due to the tension of
the magnetic field lines which maintain a more laminar flow around
the fingers of dense ejecta gas that protrude into the shocked wind
material (e.g., \citealt{2012ApJ...749..156O}). Indeed the post-shock
magnetic field is heavily modified by the fingers and the field
lines wrap around these ejecta structures, leading to a local
increase of the field strength. The interaction of the remnant with
the shell leads to a further compression of the magnetic field,
which is more effective in the NW region where the shell is the
densest. There the post-shock magnetic field reaches values
of the order of $10~\mu$G when the pre-SN field strength at 2.5 pc
was $0.2~\mu$G. As a result, the field strength is significantly
higher in the NW than in the SE region of the remnant and this
contributes in determining an asymmetry in the brightness distribution
of the nonthermal emission in the two hemispheres. Note that model
W15-IIb-sh-MHD+dec neglects the magnetic field amplification due
to back-reaction of accelerated cosmic rays, so that the enhancement
of magnetic field in the NW is purely due to the high compression
of field lines during the interaction of the remnant with the shell.
Field strengths higher by an order of magnitude (and consistent
with observations) may be reached due to magnetic field amplification.

\subsection{Reverse shock asymmetries at the age of \casa}
\label{rs_asym}

The interaction of the remnant with the asymmetric shell affects
the propagation of the reverse shock in a different way in the
eastern and western sides. This is evident from an inspection of
Fig.~\ref{prof_vsh}, which shows the profiles of the forward and reverse
shock velocities versus the position angle in the plane of the sky
at the age of \casa. For comparison, the figure also shows the
profiles derived from a model not including the interaction of the
remnant with the shell (upper left panel; model W15-2-cw-IIb-HD
presented in Paper I) and the profiles derived from the analysis
of Chandra observations of \casa\ (black and magenta diamonds;
\citealt{2022ApJ...929...57V}). At the age of \casa, the models
describing the interaction with the shell show a forward shock that
propagates with velocity between $5000$~km~s$^{-1}$ and $6000$~km~s$^{-1}$
at all position angles, thus producing results analogous to those
derived from the model without the shell and in agreement with
observations. This is a sign that the effect of the interaction of
the forward shock with the shell has ran out and, in fact, in all
the models the forward shock propagates through the $r^{-2}$ wind
density distribution\footnote{As mentioned in
Sect.~\ref{param_space}, simulations with $r\rs{sh} > 1.5$~pc (not
reported here) produce profiles of the forward shock velocity versus
the position angle in the plane of the sky with a significant
decrease in the NW side at odds with observations.}.

Conversely, the velocity of the reverse shock shows strong changes
with the position angle if the remnant interacts with the shell,
at odds with model W15-2-cw-IIb-HD that shows a reverse shock
velocity around $\approx 3000$~km~s$^{-1}$ at all position angles.
In the eastern side, where the shell is tenuous, the evolution of
the reverse shock is only marginally affected by the shell and the
shock propagates with a velocity of $\approx 3000$~km~s$^{-1}$ (as
in model W15-2-cw-IIb-HD); in the western side, where the shell
is dense, the reverse shock is slowed down by the reflected shock
driven by the interaction with the shell and, as a result, the
reverse shock appears to move inward in the observer frame as
observed in \casa. Fig.~\ref{prof_vsh} shows that the agreement
between models and observations is remarkable, producing a
reverse-shock minimum roughly where it is observed. However, we also
note that the predicted minimum is broader than observed, extending
further to the north. This suggests that the high density portion of
the modeled shell is too large (the match with the data would
improve if the high density portion of the shell subtends a smaller
solid angle as seen from the explosion center) or that the actual
shell is incomplete or irregular (and not with a regular spherical
shape as assumed).

We note that \cite{2018ApJ...853...46S} report bright nonthermal
X-ray emitting features in the interior of \casa\ due to inward-moving
shocks in the western and southern hemispheres (see also
\citealt{1995ApJ...441..307A, 1996ApJ...466..309K, 2004ApJ...613..343D,
2008ApJ...686.1094H}). These features are not reproduced by our
models which predict an inward-moving reverse shock only in the NW
hemisphere. Features similar to those observed by
\cite{2018ApJ...853...46S} might be obtained by rotating the
asymmetric shell approximately $90^{\rm o}$ clockwise about the $y$
axis. In this case, however, we found that the models do not reproduce
other asymmetries that characterize the reverse shock of \casa, in
particular the velocity profiles derived by \cite{2022ApJ...929...57V}
and the orientation of the offset between the geometric center of
the reverse shock and that of forward shock
(\citealt{2001ApJ...552L..39G}). In Appendix~\ref{new_model}, we
present an example of these models. Here we preferred to discuss
the models that most closely reproduce many (but not all) of the
reverse shock asymmetries observed in \casa. Nevertheless, our
study clearly shows that the interaction of the remnant with local
asymmetric density enhancements (as the densest portion of the shell
in our simulations) can produce inward-moving reverse shocks there.
In case of a shell with a more complex structure (e.g., more fragmented)
than modeled here, it may be possible to reproduce the locations
where inward-moving shocks are observed if density enhancements
are placed in the same locations.

\begin{figure}[!t]
  \begin{center}
    \leavevmode
        \epsfig{file=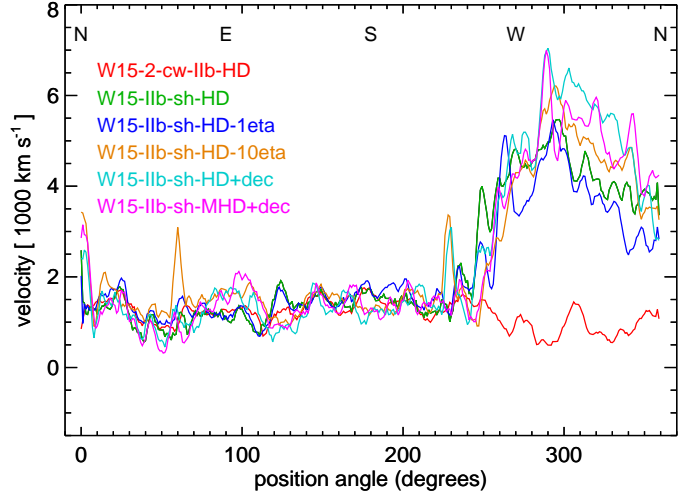, width=9cm}
	\caption{Velocity of ejecta when they enter the reverse
	shock versus the position angle at the age of \casa\ for
	the first six SNR models listed in Table~\ref{tabmod},
	including model W15-2-cw-IIb-HD presented in Paper I.}
  \label{vRS_ejframe}
\end{center} \end{figure}

The negative velocity of the reverse shock in the NW side has
important consequences on the acceleration of particles. In fact,
the ejecta enter the reverse shock with a higher relative velocity
in the western than in eastern side. Fig.~\ref{vRS_ejframe} shows
the velocities of ejecta when they enter the reverse shock at the
different position angles. The velocity is below $2000$~km~s$^{-1}$
for most of the position angles, except in the western part, where
the ejecta enter the reverse shock with a velocity between $4000$
and $7000$~km~s$^{-1}$. This implies that, only in the western part,
the reverse shock is potentially able to accelerate electrons to
high enough energies to emit X-ray synchrotron radiation: its
velocity relative to the ejecta must be well above the limit for
producing this emission ($\approx 3000$~km~s$^{-1}$;
\citealt{2020pesr.book.....V}). This may explain why most of the
X-ray synchrotron emission originates from the western part of the
reverse shock (\citealt{2008ApJ...686.1094H}) where the radio and
X-ray observations agree on the position of the reverse shock
(\citealt{2020pesr.book.....V}; see also Fig.~\ref{fig_casa}).
Interestingly, from the analysis of the 1~Ms Chandra observation
of \casa, \cite{2008ApJ...686.1094H} concluded that the dominant
X-ray synchrotron emission from the western side of \casa\ can be
justified by a local reverse shock velocity in the ejecta frame of
$\approx 6000$~km~s$^{-1}$ as opposed to a velocity $\approx
2000$~km~s$^{-1}$ elsewhere. Our models predict similar velocities.

\begin{figure*}[!t]
  \begin{center}
    \leavevmode
        \epsfig{file=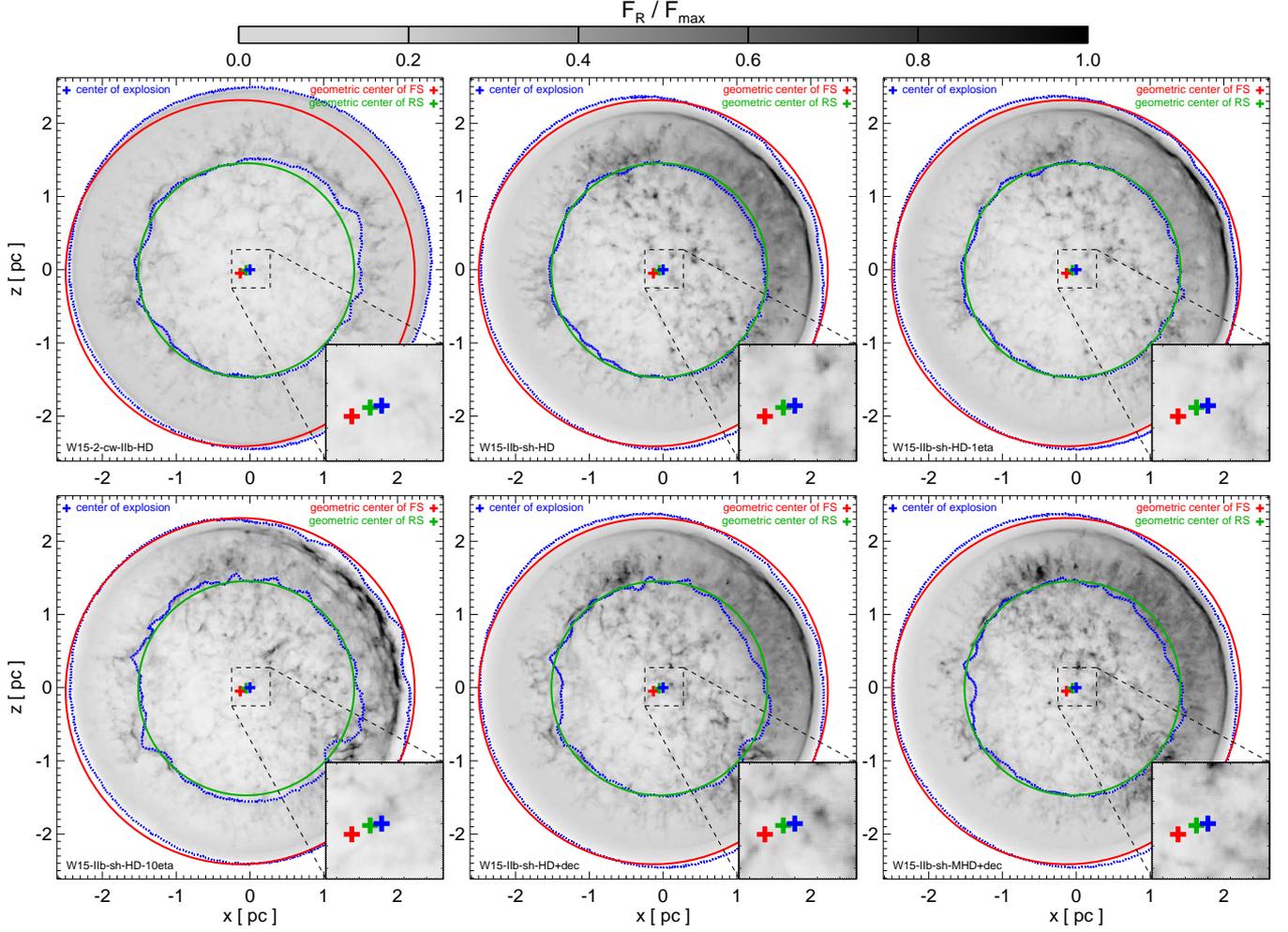, width=18.cm}
	\caption{Radio maps at the age of \casa\ synthesized
	from the first six SNR models listed in Table~\ref{tabmod},
	including model W15-2-cw-IIb-HD (presented in Paper I). The
	maps are normalized to the maximum radio flux, $F\rs{max}$,
	in model W15-IIb-sh-HD-10eta. The blue dotted contours
	show cuts of the forward and reverse shocks in the plane
	of the sky passing through the center of the explosion
	(marked with a blue cross in each panel). The red and green
	circles mark the same cuts but for spheres roughly delineating
	the forward and reverse shocks, respectively, in models
	describing the remnant-shell interaction (the circles are
	the same in all the panels); the center of these spheres
	are marked with a cross of the same color (red or green)
	in each panel. These crosses represent the geometric center
	of the forward and reverse shocks, respectively, offset to
	the SE from the center of the explosion. The inset in the
	lower right corner of each panel is a zoom of the center
	of the domain.}
  \label{fig_radio}
\end{center} \end{figure*}

The remnant-shell interaction can also alter the geometric centers
of the forward and reverse shocks. Since these shocks can be easily
traced by the nonthermal emission due to particle acceleration at
the shock fronts (e.g., \citealt{2018A&A...612A.110A}), we synthesized
the radio emission from the models at the age of \casa. The synthesis
has been performed using REMLIGHT, a code for the synthesis of
synchrotron radio, X-ray, and inverse Compton $\gamma$-ray emission
from MHD simulations (\citealt{2007A&A...470..927O, 2011A&A...526A.129O}).
Note that most of our simulations do not include an ambient magnetic
field. In these cases we synthesized the nonthermal emission assuming
a uniform randomized magnetic field with strength $1\,\mu$G in the
whole spatial domain. In model W15-IIb-sh-MHD+dec, in which the
ambient magnetic field configuration is described by the ``Parker
spiral'' (\citealt{1958ApJ...128..664P}), we synthesized the emission
by adding a background uniform randomized magnetic field with
strength $0.5\,\mu$G to the spiral-shaped magnetic field; this was
necessary to prevent a field strength very low at a distance of a
few pc from the center of explosion and to make the field strength
comparable to that assumed for the other models. We note that
non-linear amplification of the field in proximity of the shock due
to cosmic rays streaming instability is expected at the forward
shock (\citealt{2004MNRAS.353..550B}) and, most likely, the same
process is also operating at the reverse shock. Our models, however,
do not include this effect. Hence our synthetic maps cannot be
directly compared with radio observations of \casa. In fact, as
mentioned in Sect.~\ref{sec:sn-snr}, the radio emission was synthesized
as a proxy of the position of the reverse shock, which is not
expected to depend on the configuration and strength of the magnetic
field. On the other hand, we expect that higher reverse-shock
strengths may produce higher magnetic field strengths and this may
enhance the synchrotron emissivity in the NW side.

Fig.~\ref{fig_radio} shows the radio maps synthesized for the first
six SNR models listed in Table \ref{tabmod}. To better identify the
effects of the remnant-shell interaction on the structure of the
forward and reverse shocks, we compared the synthetic maps from
these models with the radio map synthesized from model W15-2-cw-IIb-HD
(upper left panel in the figure). For each model, we derived
the position of the forward and reverse shocks in the plane of the
sky (blue contours in the figure) and fitted these positions (the
contours) with circles (thus deriving the geometric centers and the
radii of the forward and reverse shocks). We found that the centers
and radii of the circles fitting the forward and reverse shocks are
very similar in models which describe the remnant-shell interaction
in Table~\ref{tabmod}. For this reason, the circles reported in
Fig.~\ref{fig_radio} (red for the forward shock and green for the
reverse shock) correspond to those derived from the fitting of the
blue contours in model W15-IIb-sh-HD-1eta and are the same in all
the panels to help in the comparison between model W15-2-cw-IIb-HD
and all the other models.

As expected the effects of the remnant-shell interaction are most
evident in the NW region, where the shell has the highest density.
In this region, both the forward and reverse shocks in models
including the shell are at smaller radii from the center of the
explosion than in model W15-2-cw-IIb-HD. This is evident from
the upper left panel in Fig.~\ref{fig_radio} where both the forward
and reverse shocks in model W15-2-cw-IIb-HD (blue contours) expand
to NW more than in the other models (red and green circles). The
forward shock slows down significantly as it propagates through the
densest portion of the shell, bringing the shock front at a smaller
distance from the center of the explosion than in model W15-2-cw-IIb-HD.
After the interaction with the shell, the forward shock starts
propagating again with the same velocity as in model W15-2-cw-IIb-HD
(see Fig.~\ref{prof_vsh}) but with a smaller radius.  As for the
reverse shock, it interacts with the reflected shock from the shell
which causes it to move inward in the observer frame in the NW
region. As a result, in models including the shell, also
the reverse shock is at smaller radii from the center of the explosion
than in model W15-2-cw-IIb-HD.

The asymmetry introduced by the interaction of the remnant with the
shell leads to the geometric centers of the forward and reverse
shocks (red and green crosses, respectively, in each panel of
Fig.~\ref{fig_radio}) that are shifted to the SE from the center
of the explosion (blue cross in the figure). This is opposite to
the result found in Paper I, where we found an offset of the geometric
centers of the two shocks toward the NW by $\approx 0.13$~pc from
the center of the explosion. In fact, in models not including the
interaction with the shell, the offset is caused by the initial
asymmetric explosion, in which most of the $^{56}$Ni and $^{44}$Ti
were ejected in the northern hemisphere away from the observer.
Therefore, in our models, the effects of the remnant-shell interaction
are opposite to those of the asymmetric explosion and dominate in
the structure of the forward and reverse shocks at the age of \casa.

Fig.~\ref{fig_radio} also shows that, at the age of \casa, the
forward shock appears more affected than the reverse shock by the
remnant-shell interaction. We note that, while the forward shock
passed through the shell at $t \approx 180$~years after the SN, the
reverse shock started to be affected by the reflected shock from
the shell at later times, namely at $t \approx 250-290$~years (see
Sect.~\ref{interaction}). This delay caused the distance between
the reverse and forward shock to gradually decrease in the time
interval between 180 and 250 years; then the distance started to increase
for $t > 250$~years when the reflected shock started to push the reverse
shock inward. At the age of \casa, the distance between the forward
and reverse shocks in the NW region is still smaller than expected
without the interaction with the shell.

An important consequence of the asymmetry introduced by the
remnant-shell interaction is that, at the age of \casa, the geometric
center of the reverse shock is offset to the NW by $\approx 0.1$~pc
from the geometric center of the forward shock (the values
range between 0.09~pc and 0.1~pc for the different models of remnant-shell
interaction). This result differs from that of Paper I in which
the geometric centers of the two shocks coincide. We note that the
offset between the two shocks caused by the remnant-shell interaction
is similar to that inferred from \casa\ observations: the latter
suggest that the reverse shock is offset to the NW by $\approx
0.2$~pc (assuming a distance of 3.4 kpc) from the geometric center
of the forward shock (e.g., \citealt{2001ApJ...552L..39G}). We note,
however, that the shocks observed in \casa\ deviate substantially
from spherical shape. In fact the geometric centers of the two
shocks were derived from azimuthal averages of the shock positions
as inferred from observations and uncertainties can be of the order
of 10 arcsec (i.e., 0.16 pc). Thus, we considered the direction of
the offset as the main feature in discerning between models.
While the difference in the values of the offset derived from
observations and from the models may not be quantitatively significant
given the uncertainties of fitting the shock locations with perfect
circles, it is encouraging that the offset predicted by our models
is consistent in extent and direction with that inferred from
observations within the uncertainties.

It is worth mentioning that models
producing, at the age of \casa, an inward-moving reverse shock in
the southern and western hemispheres (as suggested, e.g., by
\citealt{2018ApJ...853...46S}) predict an offset to the south-west
(SW) instead of NW, at odds with the observations of \casa\ (see
Appendix~\ref{new_model}). Given the idealized and simplified
description of the asymmetric shell considered in our models, the
offset develops along the direction between the center of the
explosion and the densest side of the shell. Consequently, it always
points to the region characterized by the inward motion of the
reverse shock. Interestingly, this is not the case for \casa, where
the offset points to the NW and inward-moving shocks are observed
in the southern and western hemispheres (see
\citealt{2018ApJ...853...46S}). The effects qualitatively
shown here for the NW quadrant might operate as well in response
to smaller-scale density enhancements in other directions; for
instance, an interaction of the remnant with multiple shells or a
more complex structure of the asymmetric shell may justify the
observed asymmetries.

Finally, we note in Fig.~\ref{fig_radio} that the surface brightness
of the radio emission is the highest in the NW region. This is due
to two factors: i) the post-shock plasma having the highest density
in this region as a result of the densest portion of the shell being
shocked and ii) the highest velocity of the reverse shock in the
ejecta rest frame (see Fig.~\ref{vRS_ejframe}). In model
W15-IIb-sh-MHD+dec, the compression of the magnetic field in the
interaction of the remnant with the shell further enhances the radio
emission in the NW region (see lower right panel in Fig.~\ref{fig_radio}).
We stress again that a few words of caution are needed when
comparing the radio maps derived here to actual radio observations
of \casa\ (as, for instance, the upper panel of Fig.~\ref{fig_casa}):
in fact, the synthesis of radio emission from the models does not
take into account some relevant aspects such as the unknown nature
of the ambient magnetic field configuration and its strength and
the non-linear amplification of the field in proximity of the shocks
due to cosmic rays streaming instability (\citealt{2004MNRAS.353..550B}).
In fact, our synthetic radio maps show substantial differences
with actual radio images of \casa\ (compare, for instance,
Fig.~\ref{fig_radio} with the upper panel of Fig.~\ref{fig_casa}).
Nevertheless, these maps are useful for identifying large-scale
reverse shock asymmetries caused by the remnant-shell interaction.
For instance, they predict a higher radio emission in the
western than in the eastern hemisphere of the remnant as a consequence
of the remnant-shell interaction, consistently with radio observations.

\subsection{Effects of the shell on the Doppler velocity reconstruction}
\label{dop_prj}

\begin{figure*}[!t]
  \begin{center}
    \leavevmode
        \epsfig{file=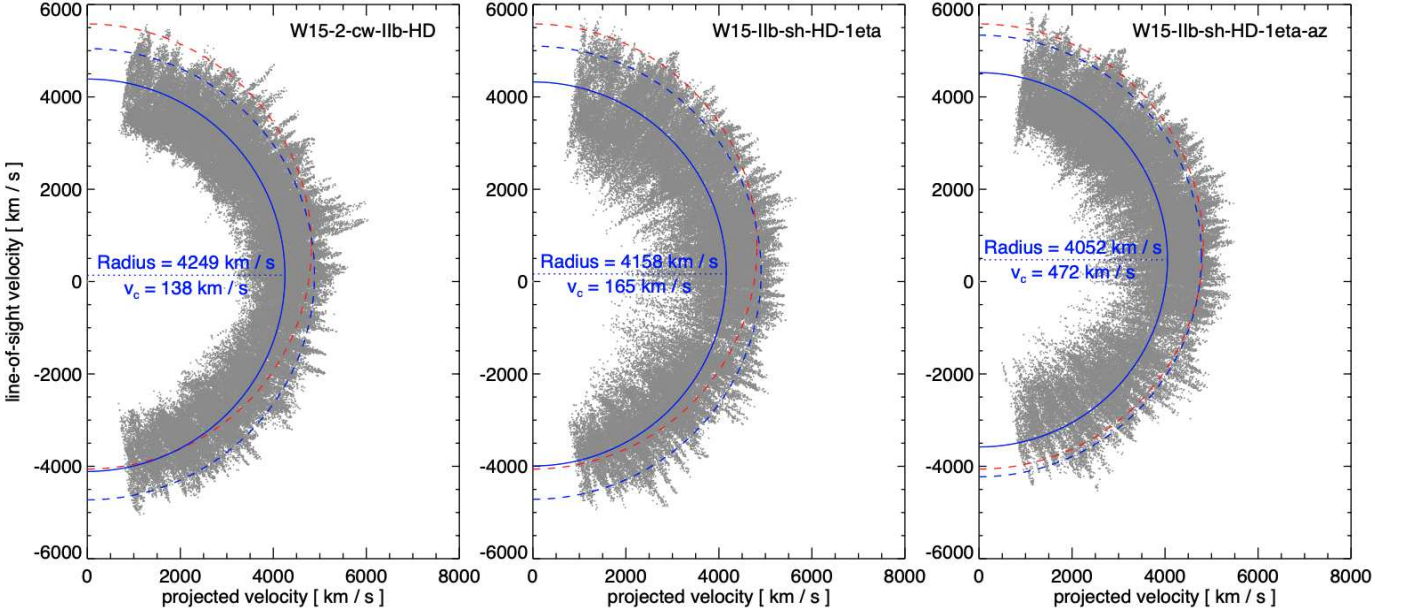, width=18.5cm}
	\caption{Projected (in the plane of the sky) and LoS
	velocities at the age of \casa\ derived from models
	W15-2-cw-IIb-HD (left panel), W15-IIb-sh-HD-1eta (center
	panel) and W15-IIb-sh-HD-1eta-az (right panel). The solid
	blue line in each panel is the best-fit semicircle to the
	data from the models; the dashed blue line shows the
	semicircle but artificially scaling the velocities to match
	the value of $v\rs{R}$ inferred from \casa\ observations;
	the dashed red line is the best-fit semicircle to actual
	data of \casa\ derived by \cite{2013ApJ...772..134M}.}
  \label{doppler_vel}
\end{center} \end{figure*}

The excellent quality of the data collected for \casa\ has allowed
some authors to perform a very accurate 3D Doppler velocity
reconstruction (\citealt{1995ApJ...440..706R, 2010ApJ...725.2038D,
2013ApJ...772..134M}) to identify possible large-scale asymmetries
of the remnant. The analysis of observed isolated knots of ejecta
showed a significant blue and redshift velocity asymmetry: the
ejecta traveling toward the observer (blueshifted) have, on average,
lower velocities than ejecta traveling away (redshifted). As a
result, the center of expansion of ejecta knots\footnote{All ejecta
knots appear to lie on a spherical shell traveling radially outward
from a unique center of expansion (e.g., \citealt{2013ApJ...772..134M}).}
appears to be redshifted with velocity $v\rs{c} = 760 \pm
100$~km~s$^{-1}$ (\citealt{2013ApJ...772..134M}). The question is
whether this asymmetry is due to the explosion dynamics (claimed
by \citealt{2010ApJ...725.2038D, 2010ApJ...725.2059I}) or to the
expansion of the remnant in inhomogeneous CSM (as suggested by
\citealt{1995ApJ...440..706R, 2013ApJ...772..134M}), possibly the
circumstellar shell investigated in the present paper.

Our simulations include both the effects of the initial large-scale
asymmetries inherited from the early phases of the SN explosion and
the effects of interaction of the remnant with an asymmetric
circumstellar shell. They therefore allow us to investigate the
possible causes of the redshift measured for the center of expansion.
From the simulations we can easily decompose the velocity of ejecta
in each cell of the spatial domain into the component projected
into the plane of the sky and the component along the LoS. The first
is the analog of the projected velocity of isolated ejecta knots
derived from their projected radii from the center of expansion in
\casa\ images and the second is the analog of the Doppler velocities
of the same knots measured from the spectra of \casa\ (e.g.,
\citealt{2013ApJ...772..134M}). For the analysis, we selected only
cells composed of at least 90\% of shocked ejecta. Hence, for each
$x$ and $z$ coordinates, we have considered the cell with the highest
kinetic energy along $y$ (i.e., along the LoS). In other words, we
selected cells characterized by a high mass of shocked ejecta and
a significant velocity.

We first checked the apparent Doppler shift of the center of expansion
introduced by the initial asymmetric SN explosion. To this end, we
considered model W15-2-cw-IIb-HD, i.e. the case of a remnant that
expands through the spherically symmetric wind of the progenitor
star without any interaction with a circumstellar shell. The left
panel of Fig.~\ref{doppler_vel} shows the LoS velocities versus the
projected velocities derived for this case. We fitted the data
points of the scatter plot with a semicircle and found that the
center of expansion is redshifted with velocity $v\rs{c} =
138$~km~s$^{-1}$. This redshift reflects the initial asymmetry of
the SN explosion, resulting in a high concentration of $^{44}$Ti
and $^{56}$Ni in the northern hemisphere, opposite to the direction
of the kick velocity of the compact object (a neutron star) pointing
south toward the observer (see \citealt{2017ApJ...842...13W}). The
radius of our best-fit semicircle corresponds to a velocity $v\rs{R}
= 4249$~km~s$^{-1}$, lower than that found by \cite{2013ApJ...772..134M}
from the analysis of observations ($v\rs{R} = 4820$~km~s$^{-1}$).
In fact, the explosion energy of our SN model ($E\rs{exp} \approx
1.5$~B; see Table~\ref{tabmod}) is a factor $\approx 1.33$ smaller
than the value inferred from the observations of \casa, $E\rs{exp}
\approx 2$~B (e.g.,~\citealt{2003ApJ...597..347L, 2003ApJ...597..362H,
2020ApJ...893...49S}). If we consider that the explosion energy was
almost entirely the kinetic energy of the ejecta, it is not
surprising that the ejecta velocities in our models are smaller
than the velocities observed in \casa. However, even by
artificially scaling the velocity of the model to match the value
of $v\rs{R}$ found in \casa, the value of $v\rs{c}$ is much lower
than that observed. In other words, an initial SN explosion
with a large-scale asymmetry, which is capable of producing a
distribution of $^{44}$Ti and $^{56}$Ni compatible with observations,
leads to a redshift of the center of expansion significantly lower
than that observed in \casa\ (compare the dashed blue and red
semicircles in the left panel of Fig.~\ref{doppler_vel}).

Then, we investigated whether an asymmetric shell as that discussed
in this paper can account for the observed redshift of the center
of expansion. The center panel of Fig.~\ref{doppler_vel} shows the
result for our reference model W15-IIb-sh-HD-1eta in which the
remnant interacts with an asymmetric circumstellar shell with the
symmetry axis perpendicular to the LoS (hence lying in the plane
of the sky; $\phi = 0$ in Eq.~\ref{eq:csm}). We found that,
in this case, the values of $v\rs{R}$ and $v\rs{c}$ are similar to
those found with model W15-2-cw-IIb-HD.  Thus, the interaction of
the remnant with a shell that is symmetric with respect to the plane
of the sky cannot contribute to determine the blue and redshift
velocity asymmetry observed in \casa. This was expected because,
in this case, the effects of the shell on the propagation of ejecta
are roughly the same in the blue and redshifted hemispheres of the
remnant.

\begin{figure}[!t]
  \begin{center}
    \leavevmode
        \epsfig{file=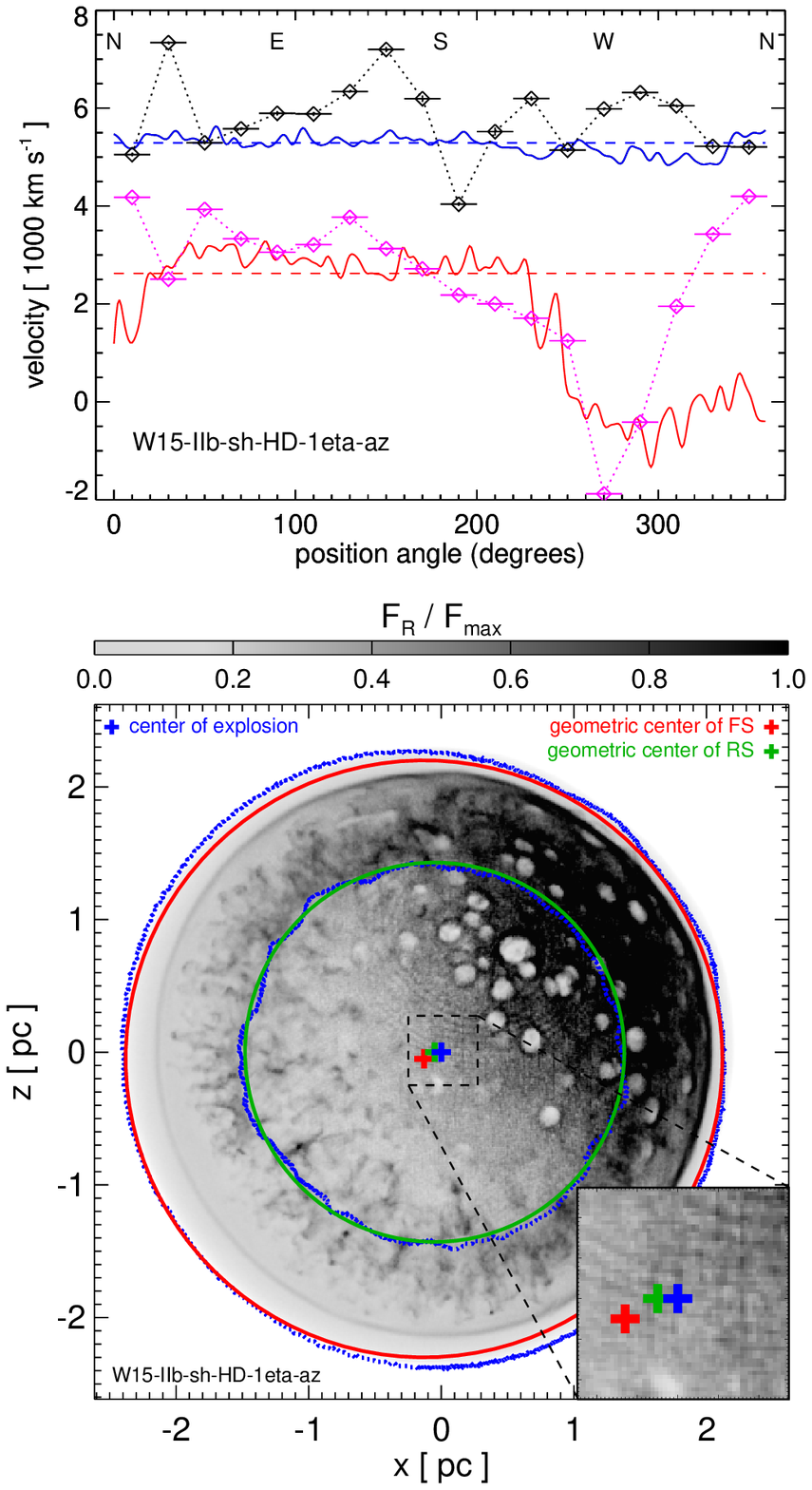, width=9cm}
        \caption{Same as in Fig.~\ref{prof_vsh} (upper panel) and
        in Fig.~\ref{fig_radio} (lower panel), but for model
        W15-IIb-sh-HD-1eta-az.}
  \label{fig_rot}
\end{center} \end{figure}

On the other hand, a denser shell on the blueshifted nearside would
inhibit the forward expansion of ejecta toward the observer, resulting
in an apparently redshifted center of expansion
(\citealt{1995ApJ...440..706R}). To test this possibility, we ran
further simulations similar to model W15-IIb-sh-HD-1eta but with
the asymmetric shell oriented in such a way that its symmetry axis
has an angle $\phi > 0$ with respect to the plane of the sky. The
right panel of Fig.~\ref{doppler_vel} shows the results for our
favorite model W15-IIb-sh-HD-1eta-az (see Table~\ref{tabmod}) in
which the shell is similar to that adopted in model W15-IIb-sh-HD-1eta
but rotated by $50^{\rm o}$ about the $z$ axis, counterclockwise
from the $[x,z]$ plane (so that the densest portion of the shell
is located in the nearside to the NW) and with a reference density
$n\rs{sh} = 20$~cm$^{-3}$ (leading to a total mass of the shell of
the order of $M\rs{sh}\approx 2\,M_{\odot}$; see parameters
in Table~\ref{tabshell}).  Increasing the reference density,
$n\rs{sh}$, with respect to the models discussed in previous sections
was necessary to keep shell densities in the western and southern
sides of the $[x,z]$ plane similar to those in model W15-IIb-sh-HD-1eta
and, therefore, to produce profiles of forward and reverse shock
velocities as those shown in Fig.~\ref{prof_vsh} (see the upper
panel of Fig.~\ref{fig_rot}) and an offset between reverse and
forward shocks of $\approx 0.1$~pc to the NW, similar to those found
in Fig.~\ref{fig_radio} (see lower panel of Fig.~\ref{fig_rot}).
With this shell configuration, the center of expansion appears to
be redshifted with velocity $v\rs{c} = 472$~km~s$^{-1}$. In this
case, an artificial scaling of the modeled velocity which matches
the observed value of $v\rs{R}$ leads to a value of $v\rs{c}$ much
closer to that inferred from observations than in the other models,
as is evident from the comparison of the dashed blue and red
semicircles in the right panel of Fig.~\ref{doppler_vel}.

By comparing the lower panel of Fig.~\ref{fig_rot} with
Fig.~\ref{fig_radio}, we note that model W15-IIb-sh-HD-1eta-az is
characterized by a significantly higher radio emission in the NW
quadrant than the other models. This is the result of the density
of the shocked shell, which is a factor of 2 higher than in the
other models. Hence, an accurate analysis of the radio emission in
the western hemisphere of \casa\ may help constraining the density
(and estimate the mass) of the putative circumstellar shell.

We conclude that the initial asymmetry in the SN explosion of \casa\
determines a redshift in the center of expansion which, however,
is not sufficient to justify the value inferred from the observations.
The encounter of the remnant with a dense circumstellar shell leads
to a further blue and redshift velocity asymmetry which makes the
apparent redshift of the center of expansion compatible with the
observations, if the nearside of the shell is denser then its
opposite side. The scenario where \casa\ interacted with an
asymmetric circumstellar shell about a hundred years ago is also
supported by the evidence that the majority of QSFs ($76$\%) have
blueshifted velocities (\citealt{1995ApJ...440..706R}), implying
that there is more CSM material in the front than in the back. This
provides additional reasons to suspect that the asymmetries associated
with the reverse shock in \casa\ can be attributed to inhomogeneities
in the surrounding material.

\section{Summary and conclusions}
\label{sec:conclusion}

In this work we investigated if some of the large-scale
asymmetries observed in the reverse shock of \casa\ (in
particular, the inward-moving reverse shock observed in the western
hemisphere of the remnant; \citealt{2022ApJ...929...57V}) can be
interpreted as signatures of a past interaction of the remnant with
a massive circumstellar shell, which is possibly a consequence of
an episodic mass loss from the progenitor star that occurred in the
latest phases of its evolution before collapse. To this end, we
performed 3D HD and MHD simulations which describe the interaction
of a SNR with an asymmetric dense shell of CSM. The SNR models are
adapted from those presented in Paper I, which describe the formation
of the remnant of a neutrino-driven SN explosion with asymmetries
and features consistent with those observed in the ejecta distribution
of \casa.

The simulations follow the evolution from the breakout of the shock
wave at the stellar surface ($\approx 20$~hours after core-collapse)
to the expansion of the remnant up to an age of $\approx 2000$~yr.
The initial conditions are provided by the output of a 3D neutrino-driven
SN simulation that produces an ejecta distribution characterized
by a large-scale asymmetry consistent with basic properties of
\casa\ (\citealt{2017ApJ...842...13W}). The interaction of the
remnant with the shell is assumed to occur during the first $\approx
300$~years of evolution, namely at an epoch prior to the age of \casa.
We explored whether back-reaction of accelerated cosmic rays,
energy deposition from radioactive decay or an ambient magnetic
field (in the absence of nonlinear amplification) have a
significant effect during the remnant-shell interaction by comparing
models calculated with these physical processes turned either on
or off. The model results at a remnant age of $\approx 350-370$~yr
were compared with the observations of \casa.

More specifically, we explored the parameter space of the
shell properties searching for a set of parameters (thickness,
radius and total mass of the shell, density contrast between the
two sides of the shell, orientation of the asymmetrical shell in
the 3D space) which is able to produce profiles of forward and
reverse shock velocities versus the position angle in the plane of
the sky similar to those observed in \casa\
(\citealt{2022ApJ...929...57V}). The analysis of the simulations
indicates the following.

\begin{itemize}
\item Initially the interaction of the remnant with the thin dense
shell slows down the forward shock (because of its propagation
through the denser medium of the shell) and drives a reflected
shock, which propagates inward and interacts with the reverse shock.
In case of an asymmetric shell with a side denser than the other,
the effects of the interaction are the largest where the shell is
the densest.

\item After the forward shock crosses the shell, it propagates again
through the $r^{-2}$ density distribution of the stellar wind with
velocities similar to those observed in models not including the
shell. In contrast, the reverse shock is highly affected by the
reflected shock from the shell and, depending on the shell density,
the reverse shock can start moving inward in the observer frame,
at odds with models not including the shell. We found that the
signatures of the past interaction of the remnant with a thin dense
shell of CSM persist in the reverse shock for a much longer time
(at least for $\approx 2000$~years in our models) than in the forward
shock (just a few tens of years).

\item Among the models analyzed, those producing reverse shock
asymmetries analogous to those observed in \casa\ predict that the
shell was thin ($\sigma \approx 0.02$~pc), with a radius $r\rs{sh}
\approx 1.5$~pc from the center of the explosion and that it was
asymmetric with the densest portion in the nearside to the NW (model
W15-IIb-sh-HD-1eta-az in Table~\ref{tabmod}).

\item In the models listed in Table~\ref{tabmod}, the remnant-shell
interaction determines the following asymmetries at the age of
\casa: i) the reverse shock moves inward in the observer frame in
the NW region, while it moves outward in other regions; ii) the
geometric center of the reverse shock is offset to the NW by $\approx
0.1$~pc from the geometric center of the forward shock and both are
offset to the SE from the center of the explosion; iii) significant
nonthermal emission is expected from the reverse shock in the NW
region because there the ejecta enter the reverse shock with a
higher relative velocity (between $4000$ and $7000$~km~s$^{-1}$)
than in other regions (below $2000$~km~s$^{-1}$).

\item The interaction of the remnant with a dense circumstellar
shell can help explain the origin of the 3D asymmetry measured by
Doppler velocities (e.g., \citealt{1995ApJ...440..706R,
2010ApJ...725.2038D, 2013ApJ...772..134M}). We found that the
asymmetry of the initial explosion, which is capable of producing
distributions of $^{44}$Ti and $^{56}$Ni remarkably similar to those
observed, leads to a redshifted center of expansion with velocity
much lower than that observed. On the other hand, the interaction
of the remnant with a shell which is denser in the (blueshifted)
nearside than in the (redshifted) farside inhibits more the forward
expansion of ejecta toward the observer, thus resulting in a center
of expansion apparently redshifted with a velocity similar to that
inferred from observations.

\item The parameters of the shell do not change significantly if
the back-reaction of accelerated cosmic rays, the energy deposition
from radioactive decay or an ambient magnetic field are taken into
account, although we have adopted a simplified modeling of
these processes.

\end{itemize}

We emphasize that the primary aim of our study was to investigate
whether the inward-moving reverse shock observed in the western
hemisphere of \casa\ may be the signature of a past interaction with
a circumstellar shell. Though our study does not aim at deriving a
unique and accurate reconstruction of the pre-SN circumstellar
shell, it clearly demonstrates that the main large-scale asymmetries
observed in the reverse shock of \casa\ can be interpreted as
evidence that the remnant interacted with a thin shell of material,
most likely ejected from the progenitor star before core-collapse.
According to our study, the shell was not spherically symmetric
but had one side denser than the other oriented to the NW and rotated
by $\approx 50^{\rm o}$ toward the observer. This caused the main
large-scale asymmetries now observed in the reverse shock of \casa.
We note that the remnant-shell interaction could also explain the
different structure of the western jet compared to the eastern one,
with the former less prominent and more jagged than the latter.
Indeed this difference may indicate some interaction of the western
jet with a dense CSM (maybe the circumstellar shell originated from
the WR phase of the progenitor star, as suggested by
\citealt{2008ApJ...686..399S}), at odds with the eastern jet which
was free to expand through a less dense environment.

According to our favorite scenario, the shell was the result of a
massive eruption that occurred between $\approx 10^4$ and
$10^5$~years before the core-collapse, if we consider that
the shell material was ejected, respectively, either at a few
$10^2$~km~s$^{-1}$ (i.e., during an hypothetic common-envelope phase
of the progenitor binary system) or at $10-20$~km~s$^{-1}$ (namely
during the red supergiant phase of the progenitor\footnote{Note
that the progenitor of \casa\ was not a red supergiant at the time
of core-collapse, namely after its outer H envelope was stripped
away (e.g., \citealt{2020NatAs...4..584K}). At the same time, we
cannot exclude that the progenitor was in the phase of red supergiant
at the time of the mass eruption that produced the circumstellar
shell.}). Then the remnant started to interact with the shell
$\approx 180$~years after the SN explosion, when the shell had an
average radius of $1.5$~pc. Some authors have found that a hypothetical
WR phase for the progenitor of \casa\ may have lasted no more than
a few thousand years, leading to a shell not larger than 1~pc (e.g.,
\citealt{2008ApJ...686..399S, 2009A&A...503..495V}). Our estimated
radius of the shell ($1.5$~pc) and the presumed epoch of mass
eruption (about $10^4-10^5$~years before the SN), therefore, suggest
that the shell has originated well before the WR phase (if any) of
the progenitor of \casa.

From our simulations, we have estimated a total mass of the shell
of the order of $M\rs{sh}\approx 2\,M_{\odot}$ (see
Sect.~\ref{dop_prj}). Considering that the progenitor of \casa\
was probably a star with a main sequence mass between 15 and
$20\,M_{\odot}$ (according to the values suggested, e.g., by
\citealt{1994AJ....107..662A, 2014ApJ...789....7L}) and that the
mass of the star before collapse was $\approx 6\,M_{\odot}$
(\citealt{1993Natur.364..507N}), we expect a total mass lost during
the latest phases of evolution of the progenitor to be between 9
and $14\,M_{\odot}$. The estimated amount of mass of the shocked
wind within the radius of the forward shock (assuming a spherically
symmetric wind with gas density proportional to $r^{-2}$) is $\approx
6\,M_{\odot}$ (see \citealt{2016ApJ...822...22O}). Thus, including
the shell, the mass of shocked CSM is $\approx 8\,M_{\odot}$, so
we can infer that less than $5\,M_{\odot}$ of CSM material is still
outside the forward shock. We note that the mass of shocked CSM
derived from our models is consistent with that inferred from the
analysis of X-ray observations of \casa\ (\citealt{1996ApJ...466..866B,
2014ApJ...789....7L}).

It is worth to mention that the earliest radio images of \casa\
collected in 1962 (i.e., when the remnant was $\approx 310$~years
old) already show a bright radio ring in which the western
region is brighter than the eastern one (\citealt{1965Natur.205.1259R}),
indicating an increase in the synchrotron emissivity there over its
value elsewhere in the ring, where our model predicts an increase
in the reverse-shock strength. A stronger reverse shock could produce
the asymmetry either through an increased efficiency of electron
acceleration or through greater magnetic-field amplification. This
explanation requires that the encounter with an asymmetric shell
must have taken place well before 1962. This is consistent with our
models listed in Table~\ref{tabmod} which predict that the remnant-shell
interaction occurred between 180 and 240 years after the SN (i.e.,
between years 1830 and 1890) and that the reflected shock from the
shell reached the reverse shock $\approx 290$~years after the SN
(i.e., in 1940). The observations collected in 1962, therefore,
could witness the early inward propagation of the reverse shock in
the western hemisphere.

Interestingly, the scenario supported by our models is consistent
with that proposed by \cite{2018ApJ...866..139K} from the analysis
of a long-exposure image centered at $1.644\,\mu$m emission collected
with the United Kingdom Infrared Telescope. From the analysis of
the spatial distribution of QSFs, these authors have found a high
concentration of these structures in the western hemisphere of
\casa\ and noted that their overall morphology is similar to that
expected from a fragmented shell disrupted by fast-moving dense
ejecta knots. Thus, they have interpreted the QSFs as the result of
interaction of the remnant with an asymmetric shell of CSM. They
have proposed that the progenitor system most likely ejected its
envelope eruptively to the west about $10^4-10^5$~years before the
explosion (consistently with our findings). \cite{2018ApJ...866..139K}
have estimated a total H+He mass of visible QSFs $\approx
0.23\,M_{\odot}$; considering a lifetime of QSFs $\gtrapprox 60$~yr,
the mass can be a little larger ($\approx 0.35\,M_{\odot}$). The
estimated mass of QSFs is lower than that estimated with our models
(of the order of $2\,M_{\odot}$). However, if the QSFs are
residuals of a shocked shell, they are certainly its densest component
and a significant fraction of the shocked material from the shell
may not be visible in the form of QSFs. Since the formation mechanism
of the QSFs is still unknown, it is hard to estimate which fraction
of the shell mass condensed to them. According to model
W15-IIb-sh-HD-1eta-az, a shell mass of the order of $2\,M_{\odot}$
is strongly favored by the Doppler velocities discussed in
Sect.~\ref{dop_prj}, so that our conclusion here is that the mass
determined for the QSFs can only be a small fraction of the total
mass of the original shell. Our models show that the radio emission
from the region of interaction of the remnant with the densest side
of the shell is sensitive to the mass density of the shell. Thus,
the analysis of radio emission may offer observational possibilities
to better constrain the shell mass.

Although our models produce reverse shock asymmetries that qualitatively
agree with the observations of \casa, some issues still remain
unexplained, for instance: i) the offset between the geometric
centers of the reverse and forward shocks ($\approx 0.1$~pc) is
much smaller than that observed ($\approx 0.2$~pc;
\citealt{2001ApJ...552L..39G}); ii) in our models this offset points
always to the region characterized by the inward motion of the
reverse shock, whilst in \casa\ the offset points to the NW but
inward-moving shocks are also observed in the southern and western
hemispheres. The discrepancies between models and observations
might originate from the idealized and simplified description of
the asymmetric shell adopted in the present study. For instance, observations
have shown some density enhancements in the north-east region
(\citealt{2020ApJ...891..116W}), where some evidence for deceleration
of the forward shock was recently reported (\citealt{2022ApJ...929...57V}).
These findings suggest that the CSM structure may be more complex
than modeled here. It is plausible that a few mass eruptions at
different epochs may have occurred before the collapse of the
progenitor star, and each ejected shell of material may have had
its own, generally non-spherical, structure. In this case, the
more complex density structure of the circumstellar shell(s) may
have induced large-scale asymmetries in the reverse shock that our
models are unable to produce. Nevertheless, our models naturally
recover reverse shock asymmetries similar to those observed, thus
supporting the scenario of interaction of the remnant with a
circumstellar shell. A more accurate description of the CSM structure
certainly require more observational input.

In the case of \casa, several lines of evidence suggest that the
progenitor star has experienced significant mass loss in its life
time. For instance, from the analysis of XMM-Newton observations,
\cite{2003A&A...398.1021W} have estimated a total mass lost from
the progenitor star before stellar death as high as $\approx
20\,M_{\odot}$ and suggested that the progenitor was a WR star that
formed a dense nebular shell before collapse. In this respect,
\cite{2009A&A...503..495V} have performed HD simulations that
describe the formation of the CSM around the progenitor of \casa\
before the SN, considering several WR life times, and the subsequent
expansion of the SNR through this CSM. Comparing the model results
with observations, they have concluded that, most likely, the
progenitor star of \casa\ did not have a WR stage or that it lasted
for less than a few thousands years (see also
\citealt{2008ApJ...686..399S}). These authors, however, considered
an almost spherically symmetric\footnote{The model includes HD
instabilities which develop at the interaction front between the
wind in the WR stage of the progenitor with the wind in the red
supergiant stage (\citealt{2009A&A...503..495V}).} cavity of the
WR wind, so that, after interaction, the reverse shock was moving
inward at all position angles. Thus, it still remains to investigate
whether an asymmetric WR wind-cavity could have similar effects to
those found here for the asymmetric shell. It is well possible that
both these structures of the CSM contributed to determine the
asymmetries observed in the reverse shock of \casa.

Erratic mass-loss episodes are known to occur in massive stars
before core-collapse. This is the case, for instance, of H-rich
massive stars that are progenitors of SNe showing evidence of
interaction with dense H- and/or He-rich CSM (hence of Type IIn
and/or Ibn, respectively), the result of mass-loss episodes occurred
shortly before core-collapse (e.g., \citealt{2010ApJ...709..856S,
2013ApJ...779L...8F, 2013ApJ...767....1P, 2014ARA&A..52..487S,
2014ApJ...789..104O} and references therein). Large eruptions are
observed in LBVs that, in fact, show large variations in both their
spectra and brightness (e.g., \citealt{1994PASP..106.1025H,
1999PASP..111.1124H}). In these cases, the dense and highly structured
CSM in which the star explodes strongly influences the dynamics of
the expanding remnant, which keeps memory of the interaction with
the dense and structured CSM for hundreds of years after the SN
(e.g., \citealt{2021A&A...654A.167U}).

In recent years, observations have shown that H-poor progenitor
stars can also experience significant mass-loss events before stellar
death. Signs of interaction of the SN blast wave with a dense medium
have been found in H-stripped Type-Ibn SNe (e.g.,
\citealt{2007ApJ...657L.105F, 2007Natur.447..829P}), Type-IIb SNe
(e.g., \citealt{2014Natur.509..471G, 2015ApJ...807...35M}) and
Type-Ib SNe (e.g., \citealt{2014ApJ...788L..14S}).
\cite{2017ApJ...835..140M} analyzed observations of SN 2014C during
the first 500 days of evolution and found the signatures of the SN
shock interaction with a dense shell of $\approx 1\,M_{\odot}$ at
a distance of $\approx 0.02$~pc, most likely the matter of a massive
eruption from the progenitor in the decades before the collapse.
Interestingly, the mass of the shell inferred from SN 2014C
observations is similar (a factor of 2 lower) to that estimated
here for the shell that interacted with \casa\ (but, in the case
of SN 2014C, the shell was ejected immediately before the collapse).
The above examples indicate mass loss events occurred in the decades
to centuries before collapse. Similar events could also have occurred
in the hundreds of thousands of years before the SN, so that the
remnant hits the relic of these mass eruptions even several hundreds
of years after the explosion. For instance, observations of the
Vela SNR have shown evidence of interaction of the remnant with a
circumstellar shell with mass $\approx 1.26\,M_{\odot}$ (again
similar to that found here) most likely blown by the progenitor
star about $10^6$~years before collapse (\citealt{2021A&A...649A..56S}).

If the scenario of a massive shell of material erupted by the
progenitor of \casa\ before collapse is confirmed, the information
on the mass of the shell and time of the episodic mass loss can be
useful to delineate the mass loss history of the progenitor star.
This information may help to shed light on the question whether the
progenitor of \casa\ was a single star or a member in a binary and,
more in general, whether the progenitor of \casa\ (and of other
Type-IIb SNe) might have lost its hydrogen envelope by an episode
of interaction with a companion star in a binary system. Most
likely the shell may be the result of one or multiple mass eruptions
from the progenitor star during the late stages of star evolution
(e.g., \citealt{2014MNRAS.438.1191S, 2014AJ....147...23L,
2014ApJ...787..163G}). However, it is interesting to note that
shell asymmetries similar to that adopted here can also be produced by
simulations describing the CSM of runaway massive stars in which
lopsided bow shock nebulae result from the wind-ISM interaction
(e.g., \citealt{2017MNRAS.464.3229M, 2020MNRAS.496.3906M,
2021MNRAS.506.5170M}). These stars, moving supersonically through
the ISM, can originate from the break-up of binary systems following
the SN explosion of one of the binary components (e.g.,
\citealt{1961BAN....15..265B, 1991AJ....102..333S, 2001A&A...365...49H,
2015MNRAS.448.3196D}) or as a consequence of dynamical multi-body
encounters in dense stellar systems (e.g., \citealt{1986ApJS...61..419G,
2003ARA&A..41...57L, 2011MNRAS.410..304G}). Ascertain the interaction
of \casa\ with an asymmetric circumstellar shell reminiscent of
those observed in runaway massive stars may help understanding the
reason why the \casa\ progenitor was stripped. Addressing the above
issues may certainly be relevant to shed some light on the still
uncertain physical mechanisms that drive mass loss in massive stars
(e.g., \citealt{2014ARA&A..52..487S}). This is of pivotal importance
given the role played by mass loss from massive stars in the galactic
ecosystem, through its influence on the life time, luminosity and
final fate of stars and its contribution on the chemical enrichment
of the interstellar medium.

\begin{acknowledgements}

We thank the anonymous referee for the careful reading of the paper.
The \PLUTO\ code is developed at the Turin Astronomical Observatory
(Italy) in collaboration with the Department of General Physics of
Turin University (Italy) and the SCAI Department of CINECA (Italy).
We acknowledge the CINECA ISCRA Award N.HP10BARP6Y for the availability
of high performance computing (HPC) resources and support at the
infrastructure Marconi based in Italy at CINECA. Additional
computations were carried out on the HPC system MEUSA at the SCAN
(Sistema di Calcolo per l'Astrofisica Numerica) facility for HPC
at INAF-Osservatorio Astronomico di Palermo. Computer resources for
this project have also been provided by the Max Planck Computing
and Data Facility (MPCDF) on the HPC systems Cobra and Draco. S.O.,
M.M., F.B. and G.P. acknowledge financial contribution from the
PRIN INAF 2019 grant ``From massive stars to supernovae and supernova
remnants: driving mass, energy and cosmic rays in our Galaxy'' and
the INAF mainstream program ``Understanding particle acceleration
in galactic sources in the CTA era''. At Garching, funding by the
European Research Council through grant ERC-AdG no. 341157-COCO2CASA
and by the Deutsche Forschungsgemeinschaft (DFG, German Research
Foundation) through Sonderforschungsbereich (Collaborative Research
Centre) SFB-1258 ``Neutrinos and Dark Matter in Astro- and Particle
Physics (NDM)'' and under Germany’s Excellence Strategy through
Cluster of Excellence ORIGINS (EXC-2094)-390783311 is acknowledged.
S.N. and M.O. thank supports from JSPS KAKENHI Grant Numbers
JP19H00693 and JP21K03545, respectively. S.N and M.O. also thank
supports from ``Pioneering Program of RIKEN for Evolution of Matter
in the Universe (r-EMU)'' and ``Interdisciplinary Theoretical and
Mathematical Sciences Program of RIKEN''. The navigable 3D graphics
have been developed in the framework of the project 3DMAP-VR
(3-Dimensional Modeling of Astrophysical Phenomena in Virtual
Reality; \citealt{2019RNAAS...3..176O}) at INAF-Osservatorio
Astronomico di Palermo.

\end{acknowledgements}

\appendix

\section{Alternative asymmetry of the circumstellar shell}
\label{new_model}

\begin{figure}[!t]
  \begin{center}
    \leavevmode
        \epsfig{file=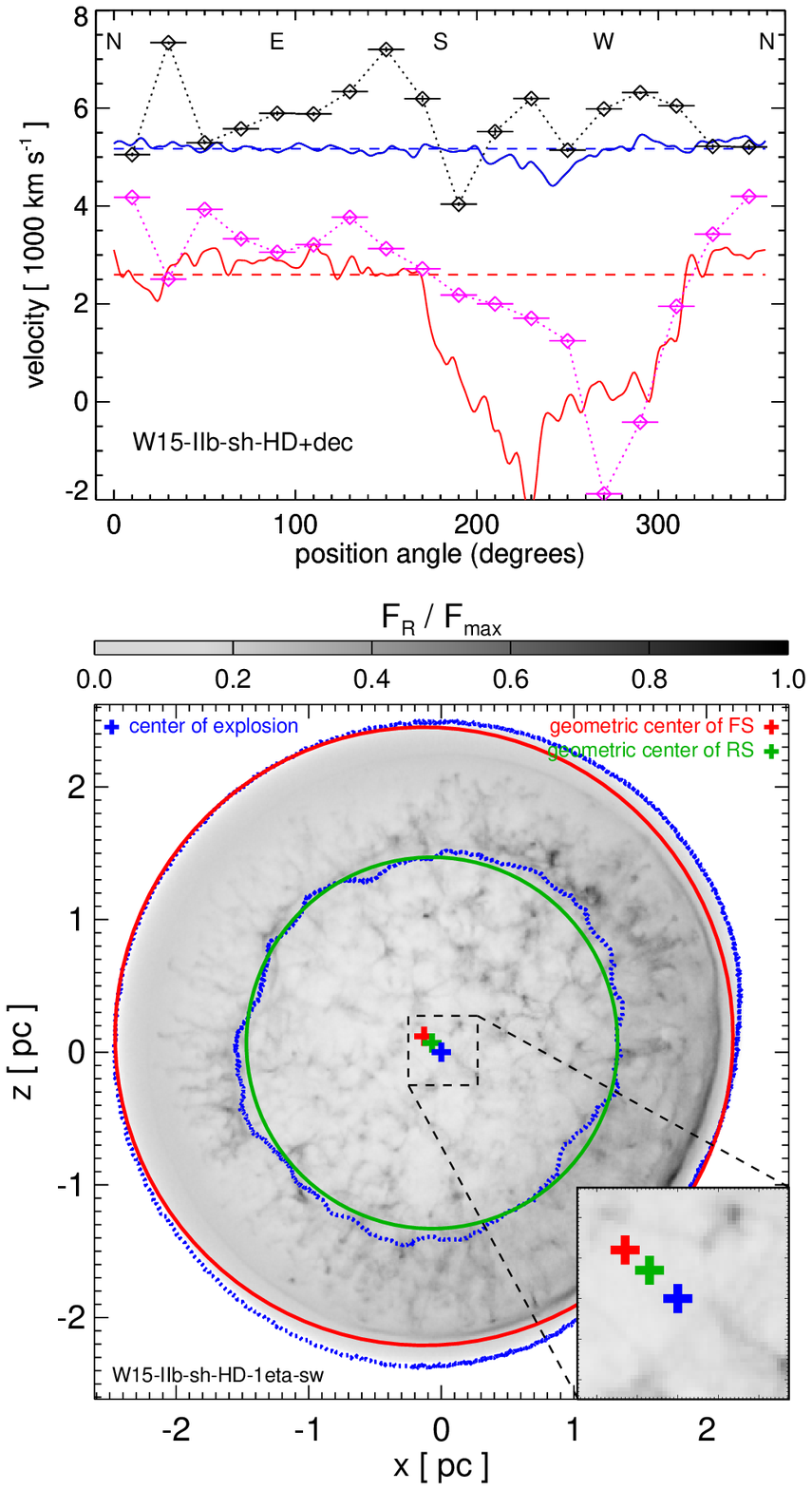, width=9cm}
	\caption{Same as in Fig.~\ref{prof_vsh} (upper panel) and
	in Fig.~\ref{fig_radio} (lower panel), but for model
	W15-IIb-sh-HD-1eta-sw.}
  \label{fig_append}
\end{center} \end{figure}

The asymmetry of the circumstellar shell investigated in the main
body of the paper leads to a reverse shock moving inward in the
observer frame, in the NW hemisphere. This feature does not fit
well with observations of \casa\ that show inward-moving shocks 
preferentially in the western and southern hemispheres (e.g.,
\citealt{1995ApJ...441..307A, 1996ApJ...466..309K, 2004ApJ...613..343D,
2008ApJ...686.1094H, 2018ApJ...853...46S}). A better match of the
models with observations may be obtained by changing the orientation
of the asymmetric shell. We considered, therefore, the setup of model
W15-IIb-sh-HD-1eta and rotated the shell by approximately $90^{\rm o}$
clockwise about the $y$ axis (model W15-IIb-sh-HD-1eta-sw). The
upper panel in Fig.~\ref{fig_append} shows the forward and reverse
shock velocities versus the position angle in the plane of the sky
at the age of \casa\ derived from the analysis of this model. As
expected, in model W15-IIb-sh-HD-1eta-sw, the reverse shock moves
inward in the western and southern hemispheres. We note, however,
that, in this way, the model is able to roughly match the velocity
profiles of forward and reverse shocks in the NW quadrant, but it
fails in reproducing the profiles in the south-west quadrant (compare
red and magenta curves in the upper panel of Fig.~\ref{fig_append}).

Furthermore, we found that other asymmetries that characterize the
remnant morphology are not reproduced by simply changing the
orientation of the asymmetric shell. In particular, model
W15-IIb-sh-HD-1eta-sw predicts an offset of $\approx 0.1$~pc to the
SW between the geometric center of the reverse shock and that of
forward shock, at odds with observations that show an offset of
$\approx 0.2$~pc to the NW (e.g., \citealt{2001ApJ...552L..39G}).
This is evident from the lower panel in Fig.~\ref{fig_append} that
reports the radio map normalized to the maximum radio flux in model
W15-IIb-sh-HD-1eta-sw. The blue dotted contours show cuts of the
forward and reverse shocks in the plane of the sky passing through
the center of the explosion (marked with a blue cross). The red and
green circles mark the same cuts but for spheres roughly delineating
the forward and reverse shocks in model W15-IIb-sh-HD-1eta-sw,
respectively. The centers of these spheres (marked with a cross of
the same color of the corresponding circles) represent the geometric
centers of the forward and reverse shocks, respectively, that are
offset to the north-east from the center of the explosion.

For the simple asymmetry considered for the circumstellar shell,
we found that, in general, the offset between the geometric center
of the reverse shock and that of forward shock points toward the
densest portion of the shell where the reverse shock propagates
inward in the observer frame. We concluded that an interaction with
multiple shells or with a shell with a more complex structure and
geometry may be required to better match the observations of \casa.

%\section{On-line material}
%\label{app:on-line}
%

\bibliographystyle{aa}
\bibliography{references}

\end{document}